\documentclass[pra,twocolumn,showpacs,amsfonts,amsmath,floatfix]{revtex4}

\usepackage[applemac]{inputenc}  
\usepackage[T1]{fontenc}       

\usepackage{amsmath}
\usepackage{amssymb}
\usepackage{dsfont}
\usepackage{bm}
\usepackage{graphicx}
\usepackage{pict2e}
\usepackage{xspace}
\usepackage{titlesec}
\usepackage[colorlinks=true,linkcolor=blue,citecolor=blue,urlcolor=blue]{hyperref}
\usepackage[dvipsnames]{xcolor}

\graphicspath{{./figures/}}

\newcommand{\bc}{\begin{center}}
\newcommand{\ec}{\end{center}}
\newcommand{\beq}{\begin{equation}}
\newcommand{\eeq}{\end{equation}}
\newcommand{\beqq}{\begin{equation*}}
\newcommand{\eeqq}{\end{equation*}}
\newcommand{\beqa}{\begin{align}}
\newcommand{\eeqa}{\end{align}}
\newcommand{\barr}{\begin{array}}
\newcommand{\earr}{\end{array}}
\newcommand{\bi}{\begin{itemize}}
\newcommand{\ei}{\end{itemize}}

\newcommand{\norm}[1]{\ensuremath{\vert\vert#1\vert\vert}}

\newcommand{\R}{\mathbb{R}}

\newcommand{\di}{\mathrm d}

\newcommand{\ket}[1]{\left\vert#1\right\rangle}
\newcommand{\bra}[1]{\left\langle#1\right\vert}
\newcommand{\lt}{\left(}
\newcommand{\rt}{\right)}
\newcommand{\lqu}{\left[}
\newcommand{\rqu}{\right]}

\input{Qcircuit}

\begin{document}

\title{Probabilistic Fault-Tolerant Universal Quantum Computation and Sampling Problems in Continuous Variables}


\author{Tom Douce$^{1}$, Damian Markham$^2$, Elham Kashefi$^{1,2}$, Peter van Loock$^3$, Giulia Ferrini$^{3,4}$}
\affiliation{$^1$ School of Informatics, University of Edinburgh, 10 Crichton Street, Edinburgh, EH8 9AB, United Kingdom \\
$^2$ Sorbonne Universit\'e, CNRS, Laboratoire d'Informatique de Paris 6, F-75005 Paris, France\\
$^3$ Institute of Physics, Johannes-Gutenberg Universit{\"a}t Mainz, Staudingerweg 7, 55128 Mainz, Germany \\
$^4$ Department of Microtechnology and Nanoscience (MC2), Chalmers University of Technology, SE-412 96 Gothenburg, Sweden
}

\date{\today}

\begin{abstract}

Continuous-Variable (CV) devices are a promising platform for demonstrating large-scale quantum information protocols. In this framework, we define a general quantum computational model based on a CV hardware. It consists of vacuum input states, a finite set of gates -- including non-Gaussian elements -- and homodyne detection. We show that this model incorporates encodings sufficient for probabilistic fault-tolerant universal quantum computing. Furthermore, we show that this model can be adapted to yield sampling problems that cannot be simulated efficiently with a classical computer, unless the polynomial hierarchy collapses. This allows us to  provide a simple paradigm for experiments to probe quantum advantage relying on Gaussian states, homodyne detection and some form of non-Gaussian evolution. We finally address the recently introduced model of Instantaneous Quantum Computing in CV, and prove that the hardness statement is robust with respect to some experimentally relevant simplifications in the definition of that model.

\end{abstract}
\maketitle 

\section{Introduction}

Continuous Variable (CV) systems are emerging as promising candidates for the implementation of quantum computation (QC) models. The main reason for this interest relies in the possibility of generating deterministically large resource states such as cluster states, composed of up to one million modes~\cite{Yoshikawa2016}. More generally, the optical CV approach includes highly efficient ways to prepare and measure sophisticated quantum states. Furthermore, new experimental techniques that are not anymore purely optical are being addressed for CV quantum information, based e.g. on microwaves resonators coupled to superconducting Josephson junctions~\cite{Wilson2011, Gu2017}, or on opto-mechanical resonators~\cite{Aspelmeyer2014, Houhou2015}. 

In contrast to their typically high efficiencies, optical CV schemes suffer from an intrinsic sensitivity to Gaussian errors such as photon loss and thermal noise. Standard approaches to quantum error correction, encoding a logical mode into many physical modes, have been proven to be inefficient when both the error itself and the operations for error correction are of Gaussian nature~\cite{Niset2009,Namiki2014}. Nonetheless, codes have been proposed that can protect logical qubits encoded into one or more optical modes against errors induced, for instance, from photon loss~\cite{Chuang1997,Wasilewski2007,Ralph2005,Bergmann2016a,Michael2016,Leghtas2013,Mirrahimi2014,Li2017,Bergmann2016b}. In particular, those schemes encoding discrete logical information into a single optical mode, such as the so-called cat codes~\cite{Leghtas2013,Mirrahimi2014,Li2017,Bergmann2016b}, are sometimes referred to as CV codes. The very first proposal of such a CV code is the famous Gottesmann-Kitaev-Preskill (GKP) encoding that later has been shown to allow for universal fault-tolerant quantum computation in the measurement-based scenario~\cite{Gottesman2001, Menicucci2014}. 
This encoding uses highly non-Gaussian states, the so-called GKP states, in order to encode DV quantum information on a CV hardware, namely onto the infinite dimensional Hilbert space of a single harmonic oscillator. These states possess a comb-like wavefunction, with peaks equally spaced placed either at even (resp. odd) multiples of $\sqrt{\pi}$ for the $0$-logical state (resp. $1$-logical state).
GKP states have also been used as ancillary input states in a recently defined sampling model in CV, namely \textsf{IQP}~\cite{Douce2017}, whose output probability distribution was shown to be hard to sample. However, these states are rather hard to produce, and only recently their experimental generation has been tackled~\cite{Fluhmann2018}.
This practical difficulty makes it desirable to define both a model of fault-tolerant CV quantum computation and sampling models that are hard to classically simulate, which do not explicitly require GKP states as ancillary states at the input of the model. 

In this work we address these two aspects. On the one hand, we define a model for universal fault-tolerant quantum computation in CV where GKP states are probabilistically generated within the model itself.
Similar to the original version of an in-principle efficient, universal and to some extent fault-tolerant linear-optics quantum computer based on probabilistic quantum gates with single-photon states, the well-known KLM model~\cite{knill2001scheme}, our model as presented here is also intended as an especially conceptual step forward rather than an immediately implementable proposal. 
Indeed, its typical success probabilities may be too low to yield an experimentally practical and scalable solution for universal quantum computation with today technology. However, it allows us to show that in principle a number of ancillary modes set in the vacuum state, a polynomial number of gates drawn from an elementary gate set and homodyne detection are sufficient for probabilistic universal fault-tolerant quantum computation.  Compared to Ref.~\cite{Lloyd1999}, we show that a finite set of gates, characterized by specific values of the gate parameters rather than a continuum, is sufficient for CV quantum computing with vacuum states.

On the other hand, we later specialize to sampling models and show that models that are hard to classically sample up to relative error can be defined, where no GKP state is needed at the input, thereby improving the results of Ref.~\cite{Douce2017} and obtaining more experimentally-friendly architectures.
These models are solely based on homodyne detection for the required measurements.

As a common ground for both applications, we provide an explicit protocol for the probabilistic generation of GKP states, by starting from vacua, a given set of elementary quantum gates and using homodyne detection. This generation method is based on the protocol of Ref.~\cite{Vasconcelos2010}, that uses input squeezed cat states, beamsplitters and homodyne measurements. We further provide an explicit decomposition of the cross-Kerr interaction necessary to generate the cat states in terms of elementary gates that belong to our models. In this way, GKP generation is subsumed within the models themselves.
As an important point, the definition of the gates in our models depends on the tolerated error probability on the computation result.

The paper is structured as follows.
In Sec.~\ref{secGKP-recall} we recall the basics of GKP encoding.
In Sec.~\ref{definition-of the models} we set the problem and define in more detail the models we are interested in. 
Section~\ref{secGKP} is dedicated to the description and characterization of the specific protocol used to generate GKP states. Then Sections~\ref{secFTUQC} and ~\ref{sec-consequences} discuss the issues of fault tolerance and its implications in the universal as well as in the subuniversal sampling models. Finally conclusions and perspectives are presented in Section~\ref{secCcl}. Throughout the paper we adopt the convention $\lqu \hat q, \hat p \rqu = i$ for the quadratures commutator, which corresponds to the relation $\hat a = (\hat q + i \hat p)/\sqrt{2}$ and fixes the vacuum fluctuations to $\Delta^2 q_0 \equiv \Delta^2 p_0 =  1/2$. 
\section{Recalling the GKP encoding}
\label{secGKP-recall}

\begin{figure}[h!]
\bc
\includegraphics[width=0.5\columnwidth]{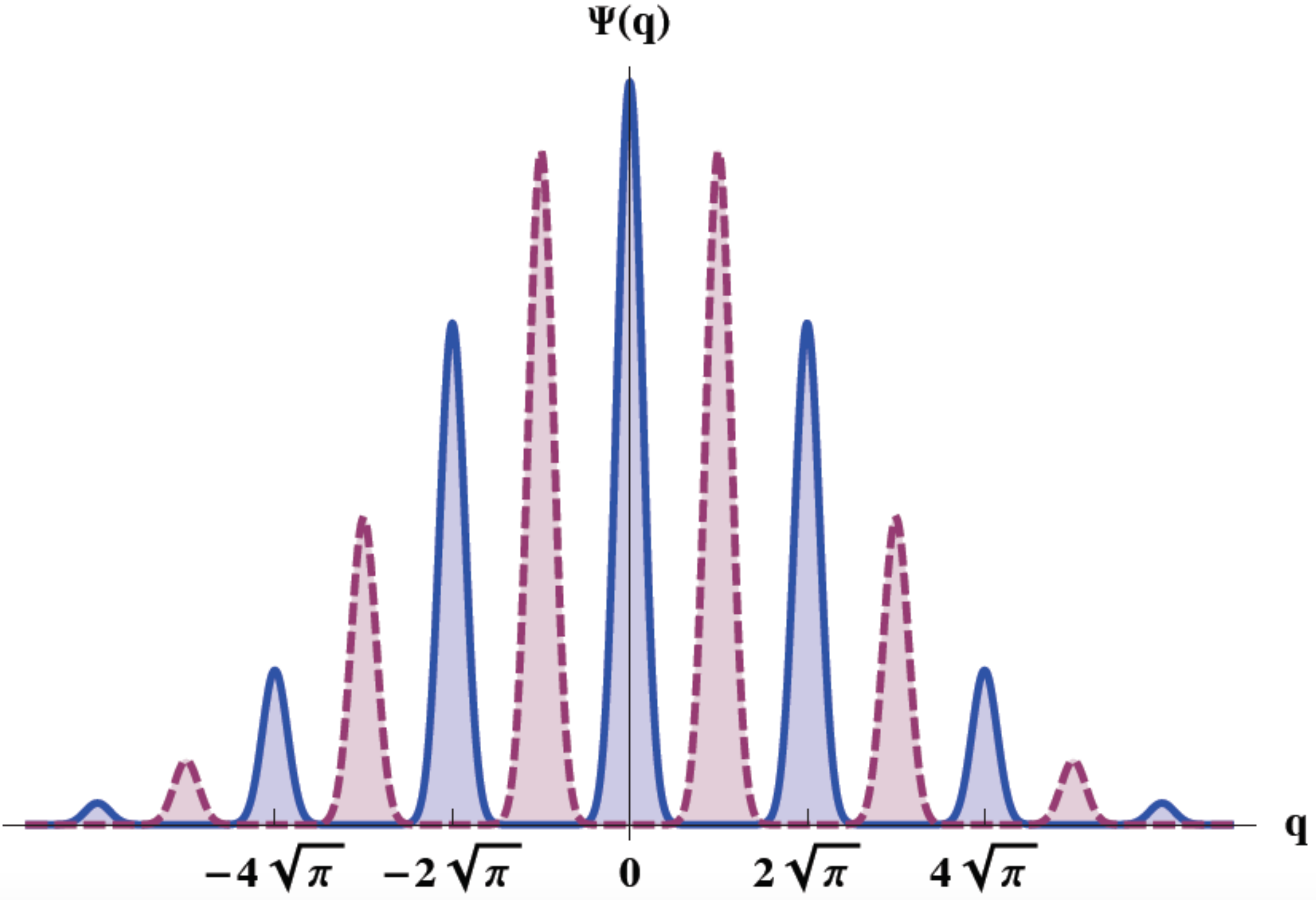}
\caption{Wavefunction in position representation of the GKP $\ket{0_G}$ state in continuous blue ($\ket{1_G}$ in dashed red) with $\delta=\sigma=0.25$ from Equation~\eqref{eqRealGKP1}.\label{fig:GKP}}
\ec
\end{figure}

GKP states are highly non-Gaussian states with a comb-like wavefunction. They allow to encode a qubit into a harmonic oscillator's Hilbert space.
Ideal GKP states, that we denote as $\ket{0_L}, \ket{1_L}$, are defined as wavefunctions made of an infinite number of Dirac peaks~\cite{Gottesman2001}:
\begin{align}\label{eqGKP}
\ket{0_L}&=\sum_n\ket{2n\sqrt\pi}_q  =\sum_n\ket{n\sqrt\pi}_p,\notag\\
\ket{1_L}&=\sum_n\ket{(2n+1)\sqrt\pi}_q  =\sum_n (-1)^{n} \ket{n\sqrt\pi}_p,
\end{align}
where $\ket{s}_q$ (resp. $\ket{t}_p$) denotes the eigenstate of the position operator of eigenvalue $s$ (resp. momentum operator of eigenvalue $t$). In the following we will omit the subscript when the situation is unambiguous.
These states realize a one-to-one correspondence between Clifford qubit operations and
Gaussian transformations. More formally, the Clifford group is mapped
to Gaussian transformations as follows:
\beq
\label{corresp1}
e^{i\sqrt\pi\hat q}\rightarrow \hat Z,\hspace{0.1cm} e^{i\hat q_1\hat q_2}\rightarrow \hat{C}_Z, \hspace{0.1cm} \hat F =e^{i\frac\pi 4 \left(\hat p^2+\hat q^2\right)}\rightarrow \hat H,
\eeq
while the non-Clifford $\hat T$ gate is mapped as
\beq
\label{corresp2}
 e^{i \frac{\pi}{4} \lqu 2 \lt \frac{\hat q}{\sqrt{\pi}} \rt^3 + \lt \frac{\hat q}{\sqrt{\pi}}\rt^2 - 2 \frac{\hat q}{\sqrt{\pi}} \rqu} \rightarrow \hat{T}.
\eeq
Realistic logical qubit states, that we indicate with $\ket{0_G}, \ket{1_G}$, are instead normalizable states, accounting for finite squeezing. Each Dirac peak is replaced by a normalized Gaussian of width $\sigma$, while the infinite sum itself becomes a Gaussian envelope function of width $\delta^{-1}$ (see Figure~\ref{fig:GKP}). Because of this and despite the fact that they are highly non-Gaussian states, we will refer to these states as to Gaussian GKP states in the following. The resulting wavefunctions are:
\small{\begin{align}\label{eqRealGKP1}
\langle q\ket{0_G}&=  \int \mathrm d u \mathrm d v\, G(u) F(v) \bra q e^{-i u \hat{p}} e^{-i v \hat{q}}  \ket{ 0 }_L  \\
&=  N_0 \sum_n\exp{\left(-\frac{(2n)^2\pi\delta^2}{2}\right)}\exp{\left(-\frac{(q-2n\sqrt\pi)^2}{2\sigma^2}\right)},\notag\\
\langle q\ket{1_G}&=  \int \mathrm d u \mathrm d v\, G(u) F(v) \bra q e^{-i u \hat{p}} e^{-i v \hat{q}} \ket{ 1 }_L \\
&= N_1 \sum_n\exp{\left(-\frac{(2n+1)^2\pi\delta^2}{2}\right)}\exp{\left(-\frac{(q-(2n+1)\sqrt\pi)^2}{2\sigma^2}\right)},\notag
\end{align}}
where we have introduced the Gaussian noise distributions
\beq
G(u) = \frac{1}{\sigma \sqrt{2 \pi}} e^{-\frac{u^2}{2 \sigma^2}}; \hspace{0.5cm} F(v) = \frac{1}{\delta \sqrt{2 \pi}} e^{-\frac{v^2}{2 \delta^2}},
\eeq
and $N_0$ and $N_1$ are normalization constants.
In the following, for conceptual clarity as well as simplicity, we consider symmetric GKP states, which have symmetric noise properties in the two quadratures, and are characterized by $\sigma = \delta$.

Ancillary $\ket{0_G}$ GKP states serve as resources to achieve fault tolerance in CV~\cite{Gottesman2001, Menicucci2014}. The idea in~\cite{Gottesman2001} is to entangle the state to be corrected at a given step of the computation with an ancillary GKP state, and then measure the ancillary modes by means of homodyne detection. One can show that in this way the noise in the $\hat q$ quadrature of a GKP encoded quantum state can be replaced by the noise of the ancillary $\ket{0_G}$ state. Repeating this gadget after a Fourier transform allows for correction of the other quadrature, thereby keeping the error below a desired amount. In Sec.~\ref{secFTUQC} we address in detail how this error correction procedure can be used in order to ensure fault tolerance in our model.

\section{Definition of the models}
\label{definition-of the models}

In this section we define the quantum computational models we are interested in. 
We first describe the CV universal quantum computational model. Then, we turn to subuniversal models and define the corresponding sampling problems.

\subsection{Probabilistic universal fault-tolerant quantum computation in CV}
\label{definition-of the modelsA}

Our model for probabilistic, universal and fault-tolerant quantum computing in CV is based upon the following elements:

(i) the multimode input state is initialized in the vacuum $\ket0^{\otimes m}$, where $m$ is the total number of modes;

(ii) the gates composing the circuit are drawn from the following finite sets: 
\begin{align}
\label{gates}
A_1&=\left\{e^{id\hat q},e^{is\hat q^2},e^{ic\hat q^3},e^{ib\hat q_1\hat q_2}\right\},\\
\label{gates2}
A_2&=\left\{  e^{i \hat{q} \sqrt{\pi}}, e^{i\frac\pi 4 \left(\hat p^2+\hat q^2\right)},  e^{i \hat{q}_1 \hat{q}_2},   e^{i \frac{\pi}{4} \left[ 2 \left( \frac{\hat q}{\sqrt{\pi}} \right)^3 + \left( \frac{\hat q}{\sqrt{\pi}}\right)^2 - 2 \frac{\hat q}{\sqrt{\pi}} \right]} \right \},  
\end{align}
where the parameters in $A_1$ will be fixed later. They will be determined by the desired precision on the computation result. In contrast to Ref.~\cite{Lloyd1999}, the CV gates in Eqs.(\ref{gates}) and (\ref{gates2}) are characterized by specific values of the gate parameters, instead of spanning the full real axis.  
The linear and quadratic gates in Eqs.(\ref{gates}) and (\ref{gates2}) are generally regarded as experimentally feasible. Note that $e^{i\frac\pi 4 \left(\hat p^2+\hat q^2\right)}\equiv\hat F$ is simply the Fourier transform. A non-Gaussian, experimentally challenging gate has also been included to  $A_1$ (with powers greater than 2 in $\hat q$ and/or $\hat p$ ), and a similar non-Gaussian gate appears in  $A_2$;

(iii) the measurements are done via homodyne detection in the momentum quadrature, i.e. by measuring $\hat p$, which corresponds to approximately measuring in the GKP basis $\{\ket{+/-_G}\}$~\cite{Gottesman2001}. The homodyne detection may have a finite resolution, which is modeled by the finitely-resolved $\hat p^{\eta}$ operator defined as~\cite{Paris2003, Douce2017}
\beq
\label{operator-proj}
\hat{p}^{\eta} = \sum_{k = - \infty}^{\infty} p_k \int_{- \infty}^{\infty} \di p\, \chi^\eta_k(p) \vert p \rangle \langle p \vert \equiv  \sum_{k = - \infty}^{\infty} p_k \hat{P}_k
\eeq
with $\chi^\eta_k(p) = 1$ for $p \in [p_k - \eta, p_k + \eta]$ and $0$ outside, $p_k = 2 \eta k$ and $2 \eta$ the resolution, associated with the width of the detector pixels. As a technical remark, note that, provided that we can find an integer $K$ such that $\sqrt\pi=K\eta$, this binning is  consistent with the dichotomy at the level of logical measurements. 

We prove that this model is at least as powerful as the standard qubit-based quantum computers by showing that any \textsf{BQP} \footnote{For Bounded Quantum Polynomial time, the class corresponding to the problems we believe could be efficiently solved by a quantum computer.} instance decided by a quantum circuit working with qubits can be mapped to a probabilistic CV circuit with a constant overhead. As mentioned above, the mapping relies on the ability to encode qubits in a quantum harmonic oscillator through the GKP procedure. More specifically we first have to generate (probabilistically) the GKP qubits in CV by applying the gate set $A_1$ and the Fourier transform upon several vacuum states. Then we use the gate set $A_2$: it is the exact analog of the universal DV gate set in the subspace spanned by the GKP states $\{\ket{0/1_G}\}$ -- as presented in Eqs.(\ref{corresp1}) and (\ref{corresp2}). Finally the noise issue can be addressed using a combination of gates following the discussion in Sec.~\ref{secFTUQC}.

\subsection{Subuniversal models and sampling}
\label{definition-of the modelsB}

Beyond fault-tolerant universal quantum computation, we address two subuniversal models of quantum computation that are associated to two respective sampling problems.

\subsubsection{CV random circuit sampling}\label{secSub1def}

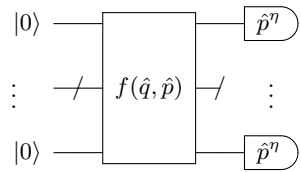
\begin{figure}[h]
\begin{minipage}{.3\columnwidth}
\centering
$$
\Qcircuit @C=1.em @R=1.5em {
\lstick{\ket0}   & \qw & \multigate{2}{f(\hat q, \hat p) } & \qw & \measureD{\hat{p}^{\eta}} \\
\lstick{\vdots\ \ \ }  & {/}\qw  & \ghost{f(\hat q, \hat p) } &  {/}\qw & \vdots\\
\lstick{\ket0} & \qw & \ghost{f(\hat q, \hat p)} &  \qw & \measureD{\hat{p}^{\eta}}
}
$$
\end{minipage} 
\caption{\label{figcirc} Sampling architecture with Gaussian input state and homodyne detection. The finitely-resolved homodyne measurement $\hat{p}^{\eta}$ has resolution $2 \eta$.}
\end{figure}

The first subuniversal model that we consider is illustrated in Fig.\ref{figcirc}. 
This family of circuits share the same elementary gates as the universal model defined in Sec.\ref{definition-of the modelsA}, namely the gate sets $A_1$ in Eq.\eqref{gates} and $A_2$ in Eq.\eqref{gates2}.

In analogy to the family of circuits with random gates drawn from an elementary set for qubits, we refer to this architecture as to CV random circuit sampling~\cite{Bouland2018}. 
Finite resolution in the homodyne detection ensures that we can associate well-defined probabilities to the continuous measurement outcomes through binning. Note that the detection modeled in Eq.(\ref{operator-proj}) is equivalent to perfectly resolved homodyne detectors followed by a discretization (binning) of the measurement outcomes. 

This model is not universal anymore, since the randomness occurring at the level of the measurement is not counteracted by post-selection. Nevertheless, we show that exact sampling -- or equivalently sampling up to relative error -- from the probability distribution of the measurement outcomes of the circuit family shown in Fig.\ref{figcirc} is classically hard, in the worst case scenario.

\subsubsection{CV Instantaneous Quantum Computing with input squeezed states}\label{secSub2def}

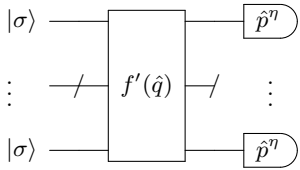
\begin{figure}[h]
\begin{minipage}{.69\columnwidth}
\centering
$$
\Qcircuit @C=1.2em @R=1.5em {
\lstick{\ket{\sigma}}   & \qw & \multigate{2}{f'(\hat q)} & \qw & \measureD{\hat{p}^{\eta}} \\
\lstick{\vdots\ \ \ }  & {/}\qw  & \ghost{f'(\hat q)} &  {/}\qw & \vdots\\
\lstick{\ket{\sigma}} & \qw & \ghost{f'(\hat q)} &  \qw & \measureD{\hat{p}^{\eta}}
}
$$
\end{minipage}
\caption{\label{figCVIQP} \textsf{IQP} circuit in CVs. $\ket{\sigma}$ are finitely squeezed states with variance $\sigma$ in the $\hat p$ representation. The gate $f'(\hat q)$ is a uniform combination of elementary gates from the set in Eqs.(\ref{eq:gates-sub2}) and (\ref{eq:gates-sub2bis}). The finitely-resolved homodyne measurement $\hat{p}^{\eta}$ has resolution $2 \eta$. }
\end{figure}

The second subuniversal model we are interested in is Instantaneous Quantum Computing in CV (Fig.\ref{figCVIQP}). 

This model is composed of the same elementary gates as those of the model in Fig.\ref{figcirc}, except for the Fourier transform, namely 
\begin{align}
\label{eq:gates-sub2}
\tilde A_1 &= \left\{e^{i \tilde d\hat{q} }, \,   e^{i \tilde s \hat{q}^2 }, e^{i\tilde c \hat{q}^3}, \,  e^{i \tilde b\hat{q}_1 \hat{q}_2} \right \},\\
\label{eq:gates-sub2bis}
\tilde A_2&=\left\{  e^{i \hat{q} \sqrt{\pi}},  e^{i \hat{q}_1 \hat{q}_2},   e^{i \frac{\pi}{4} \left[ 2 \left( \frac{\hat q}{\sqrt{\pi}} \right)^3 + \left( \frac{\hat q}{\sqrt{\pi}}\right)^2 - 2 \frac{\hat q}{\sqrt{\pi}} \right]} \right \}.
\end{align}
Therefore, all the gates in this model are diagonal in the position representation. We require momentum squeezed states ${\ket{\sigma}}$  with $\sigma<1$ to be present at the input.
This model is a simpler version than the one introduced in Ref.~\cite{Douce2017}. Specifically we show here that (i) we do not need GKP states as resource states and (ii) the squeezing parameter in the input states is constant and does not depend on the circuit size. 
These features make the present model more experimentally realistic, and yet we will prove that it retains its classical hardness (again for the exact probability distribution, and in the worst-case scenario).

The proofs of computational hardness for both sampling models, as well as the universality of the computational model presented in Sec.\ref{definition-of the modelsA}, will be based on the ability to synthesize GKP states by means of a sequence of the elementary gates that belong to the models themselves. To this end, in the following section we show how the approximate GKP states generation can be decomposed in terms of elementary gates.

\section{Approximate GKP states from a finite set of CV gates}\label{secGKP}

The main result we prove in this section is that there exists a finite number of CV gates that combined together allow one to generate approximate GKP states using vacuum input states and homodyne detection. 

Our protocol for GKP generation can be divided in two steps: 1) probabilistic generation of cat states from vacua; 2) probabilistic generation of approximate GKP states from cat states. The former protocol is based on that of  Ref.~\cite{Vitali1997} and on gate decomposition~\cite{Sefi2011}, while the latter is based on Ref.~\cite{Vasconcelos2010}. We detail here below first step 2 and later step 1, focussing on the respective associated fidelities and success probabilities.
The following subsections are rather technical and the uninterested reader may skip them and move to Sec.\ref{secFTUQC} where we take advantage of this protocol.

\subsection{GKP states from cat states}

\begin{figure}[h]
\bc
\includegraphics[width=0.8\columnwidth]{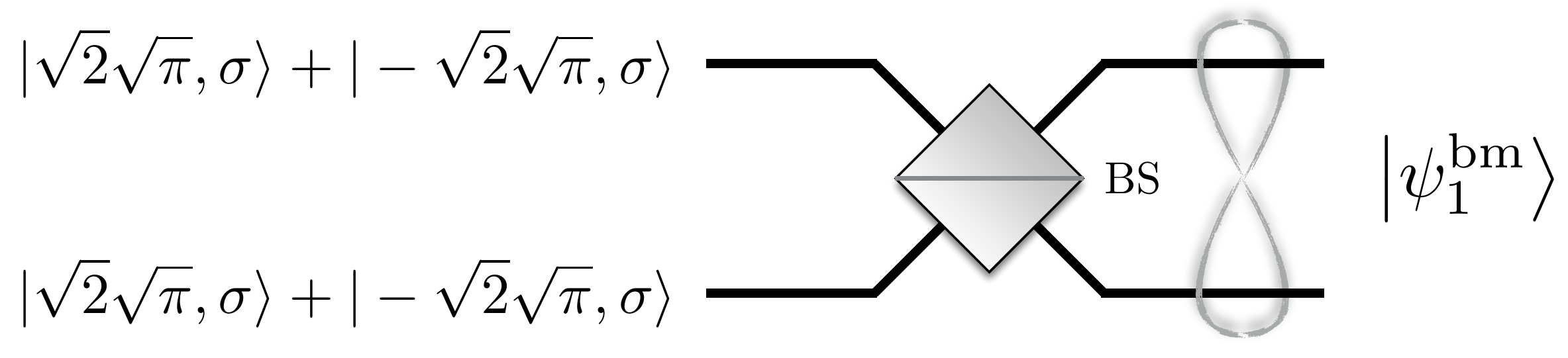}
\caption{First iteration of the protocol of Ref.~\cite{Vasconcelos2010} for the probabilistic generation of binomial GKP states: two cat states are squeezed, then combined at a beamsplitter. The $\hat p$ quadrature is then measured in one of the output modes. \label{figGlancyCat}}
\ec
\end{figure}

The protocol designed by Vasconcelos and co-authors in~\cite{Vasconcelos2010} relies on squeezed cat states combined in a linear optical network and measured by homodyne detection. The basic idea is (i) prepare two cat states $\ket\alpha+\ket{-\alpha}$ of real amplitude $\alpha$ (ii) squeeze them to reproduce the squeezing of the peaks in a GKP state (iii) send them to a balanced BS (iv) measure the $\hat p$ quadrature of one the modes. If the outcome of the homodyne detection is 0, the output state contains three Gaussian peaks enveloped by a binomial distribution, i.e. it is a {\it binomial state},  that approximates a GKP state. For this reason, and in order to avoid confusion with the binomial codes introduced in Ref.~\cite{Michael2016}, we will refer to these states as to binomial GKP states in what follows. The first iteration of this scheme is illustrated in Fig.~\ref{figGlancyCat}.
This scheme can then be repeated to produce higher order binomial GKP states possessing a larger number of peaks, better approximating Gaussian GKP states. More specifically, the $m$th binomial GKP state's position wavefunction reads:
\beq\label{eqBinom}
_q\langle q\ket{0_m}=\frac{\pi^{-1/4}}{\sqrt{\binom{2^{m+1}}{2^m}\sigma}}\sum_{i=0}^{2^m} \binom{2^m}{i} e^{-\frac{(q-2\sqrt\pi(i-2^{m-1}))^2}{2\sigma^2}},
\eeq
where $\sigma$ characterizes the squeezing of each peak. 
In general, let $m$ be a positive integer. We set the amplitude of the cat states to $\alpha_m=\sqrt2^{m-1}\sqrt\pi\sigma^{-1}$, where $\sigma$ corresponds to the amount of squeezing of the individual peaks in the GKP position wavefunction. 

\subsubsection{Quantification of the quality of the binomial GKP states}

We compare the binomial GKP states obtained through this procedure to the standard Gaussian GKP states  in Eq.\eqref{eqGKP}. We stress that in a binomial GKP state, although the individual peaks are standard Gaussians, the envelope is described by binomial coefficients. More specifically, the weight of the $i$-th peak, in terms of probabilities, is given by 
\beq\label{eqBinom}
\frac1{\binom{2^{m+1}}{2^m}} \binom{2^{m}}{i}^2
\eeq
instead of the Gaussian function characterized by a squeezing parameter given in Eq.(\ref{eqRealGKP1}). We aim at finding the closest Gaussian function approximating the distribution in Eq.\eqref{eqBinom}. Using the central limit theorem in the limit of large $m$, the binomial distribution of parameters $2^m$ and $1/2$ leads to a normal distribution of mean $2^{m-1}$ and variance $2^{m-2}$. Mathematically it means we have the following relation:
\beq\label{eqLGN}
\frac1{2^{2^m}} \binom{2^{m}}{i}\approx \frac1{\sqrt{2\pi2^{m-2}}}\exp\left(-\frac{(2^{m-1}-i)^2}{2\cdot2^{m-2}}\right).
\eeq
So necessarily the distribution in Eq.\eqref{eqBinom}, which corresponds to the left-hand-side squared, will be associated with a Gaussian of mean $2^{m-1}$ and variance $2^{m-3}$. Taking into account the proper normalisation we obtain:
\beq\label{eqGaussApprox}
\frac1{\binom{2^{m+1}}{2^m}} \binom{2^{m}}{i}^2\approx\frac1{\sqrt{2\pi2^{m-3}}}\exp\left(-\frac{(2^{m-1}-i)^2}{2\cdot2^{m-3}}\right).
\eeq
In a standard, Gaussian enveloped GKP state, recall from Eq.\eqref{eqRealGKP1} that the weight reads $\exp(-(2j\sqrt\pi)^2\sigma^2/2)$ for the $j$-th peak of the wavefunction. Using Eq.\eqref{eqLGN} it yields for the corresponding squeezing parameter $\sigma$:
\beq
\label{eq:m-vs-sigma}
\sigma^2=\frac1{2^{m}\pi}.
\eeq
We stress that due to Eq.(\ref{eq:m-vs-sigma}) the effective squeezing in the binomial GKP states depends on  the number of iterations of the protocol: the higher the number of iterations, the higher the squeezing. 
Thus we have identified the closest Gaussian GKP state to the binomial GKP state of Eq.\eqref{eqBinom}. Its position wavefunction reads:
\begin{align}
_q\langle q\ket{0_G}=\frac{(2^{m-2})^{-1/4}}{\sqrt{\sigma\pi}}\sum_{j=-\infty}^{+\infty}e^{-\frac{(2j\sqrt\pi)^2}{2\cdot2^{m}\pi}}e^{-\frac{(q-2j\sqrt\pi)^2}{2\sigma^2}}.
\end{align}
We may now compute the fidelity between the binomial GKP state produced through the protocol of Ref.~\cite{Vasconcelos2010} and the Gaussian GKP state. To simplify the calculations we assume that the Gaussian peaks are narrow enough to be considered as orthogonal. The inner product thus reads:
\begin{align}
\vert\langle 0_m\ket{0_G}\vert&\approx\frac{\pi^{-1/4}\left(2^{m-2}\right)^{-1/4}}{\sqrt{\binom{2^{m+1}}{2^m}}}\sum_{i=0}^{2^m}\sum_{j=-\infty}^{+\infty}\binom{2^m}{i}e^{-\frac{(2j\sqrt\pi)^2}{2\cdot2^{m}\pi}}\notag\\
& \int\di q\frac{1}{\sqrt\pi\sigma}e^{-\frac{(q-2\sqrt\pi(i-2^{m-1}))^2}{2\sigma^2}}e^{-\frac{(q-2j\sqrt\pi)^2}{2\sigma^2}}.
\end{align}
This assumption also implies that the integral vanishes except for the $2^m+1$ cases when $j=i-2^{m-1}$, and it is then properly normalized. So we simply have to focus on the overlap of the envelopes, namely:
\beq\label{eqOverlap}
\vert\langle 0_m\ket{0_G}\vert\approx\frac{\pi^{-1/4}\left(2^{m-2}\right)^{-1/4}}{\sqrt{\binom{2^{m+1}}{2^m}}}\sum_{i=0}^{2^m}\binom{2^m}{i}e^{-\frac{(i-2^{m-1})^2}{2^{m-1}}}.
\eeq
Table~\ref{tabGKPBinom} summarizes the results for different values of the iteration parameter $m$. Note that the fidelity between binomial and Gaussian GKP states becomes quite large already for few iterations.

\subsubsection{Finite resolution and success probability}

The protocol for GKP states synthesis from cat states that we have described above is probabilistic, and indeed it works only if all $2^m-1$ homodyne detections yield outcome 0. 
In order for the success probability not to vanish completely, homodyne detection must have finite resolution. 
Hence, we introduce a resolution $\eta$ for the homodyne detection as a binning of the real axis, consistently with Eq.(\ref{operator-proj}). The projector on outcome zero associated with all homodyne detections thus reads:
\beq
P_0^m=\int_{-\eta}^\eta\di{\bf s}\ket{\bf s}_p\bra{\bf s},
\eeq
where ${\bf s}$ is a $2^m-1$ dimensional real vector and the integration is over the box $[-\eta,\eta]^{2^m-1}$. 
In the following we quantify the success probability associated with GKP state synthesis using this detection binning.

We first focus on the $2^m$-mode quantum state $\ket{\psi_m^{\rm bm}}$ right before the measurements are performed.
We denote $\ket{\alpha, \sigma}$ a displaced squeezed vacuum state, where $\alpha\in\R$ corresponds to the displacement and $\sigma$ to its squeezing. The position wavefunction of such a state is:
\beq\label{eqDispSq}
_q\langle q\ket{\alpha,\sigma}=\frac{1}{\pi^{1/4}\sigma^{1/2}}\int\di q\,e^{-\frac{(q-\sqrt2\alpha)^2}{2\sigma^2}}.
\eeq
We consider as input state the squeezed cat state. With these notations, the input state $\ket{\psi^{\rm in}_m}$ to generate the $m$th binomial GKP state is merely:
\beq
\ket{\psi^{\rm in}_m}=\frac1{\sqrt N}\bigotimes_{i=1}^{2^m}\left(\ket{\alpha_m,\sigma}_i+\ket{-\alpha_m,\sigma}_i\right),
\eeq
where the subscript $i$ labels the mode, $\alpha_m=\sqrt2^{m}\sqrt\pi$ and $N$ is the state norm. Note that, in this simple case, the action of a beamsplitter on two identically squeezed states reads:
\beq\label{eqBS}
\ket{\mu,\sigma}\ket{\lambda,\sigma}\overset{\rm BS}\longmapsto\ket{\frac{\mu-\lambda}{\sqrt2},\sigma}\ket{\frac{\mu+\lambda}{\sqrt2},\sigma}.
\eeq
So the global state after the beamsplitters is a sum of many $2^m$-fold tensorial products involving displaced squeezed states. We can actually isolate the output mode and using combinatorial arguments finally write down $\ket{\psi_m^{\rm bm}}$:
\begin{align}\label{eqPsibm}
\ket{\psi_m^{\rm bm}}&=\frac1{\sqrt N}\sum_{i=-2^{m-1}}^{2^{m-1}}\ket{2i\sqrt\pi,\sigma}\notag\\
&\sum_{j=1}^{\binom{2^m}{2^{m-1}+i}}\ket{\alpha_1^{i,j},\sigma}\ldots\ket{\alpha_{2^m-1}^{i,j},\sigma},
\end{align}
where the $\alpha_k^{i,j}$'s are multiples of $\sqrt\pi$ and bounded in absolute value by $2^{m-1}\sqrt\pi$.


Now recall that the momentum wavefunction of a displaced squeezed vacuum state of real amplitude $\alpha$ reads: 
\beq\label{eqProjBeta}
_p\langle s\ket{\alpha,\sigma}=\frac{\sqrt\sigma}{\pi^{1/4}}e^{-i\sqrt2\alpha s}e^{-\frac{s^2\sigma^2}2}.
\eeq
Let $\rho_r$ be the density matrix obtained after tracing out the unmeasured mode. Starting from Eq.\eqref{eqPsibm}, we have (note the relabeling of the variables):
\begin{align}
\rho_r=\frac1N\sum_{i=0}^{2^m}&\sum_{j,j'=1}^{\binom{2^m}{i}}\ket{\alpha_1^{i,j},\sigma}\ldots\ket{\alpha_{2^m-1}^{i,j},\sigma}\notag\\
&\bra{\alpha_1^{i,j'},\sigma}\ldots\bra{\alpha_{2^m-1}^{i,j'},\sigma}.
\end{align}
We may now derive the success probability of the protocol, i.e. the probability that $2^m-1$ homodyne detections yield outcome 0. In the state before the measurement, the maximum (absolute) value of all amplitudes is $2^{m-1} \sqrt\pi $. So if $\eta\cdot2^{m-1}\sqrt\pi\ll1$ and $\eta\cdot\sigma_m\ll1$ (and only the first assumption is sufficient if  Eq.\eqref{eq:m-vs-sigma} holds) we have from Eq.\eqref{eqProjBeta}:
\beq\label{eqProj}
_p\langle s\ket{\alpha_k^{l,l'},\sigma_m}\approx\frac{\sqrt{\sigma_m}}{\pi^{1/4}},
\eeq
for all triples $(k,l,l')$ and all $s\in[-\eta,\eta]$. The probability of hitting 0 for all measurements, $p_m(0)$, is then given by
\begin{align}
p_m(0)&={\rm Tr}\left(P_m(0)\rho_r\right)=\int_{-\eta}^{\eta}\di{\bf s}\bra{\bf s}\rho_r\ket{\bf s}.
\end{align}
Using Eq.\eqref{eqProj} we get:
\begin{align}\label{eqSucProba}
p_m(0)&=\frac1N\int_{-\eta}^{\eta}\di{\bf s}\sum_{i=0}^{2^m}\sum_{j,j'=1}^{\binom{2^m}{i}}\left\vert\left(\frac{\sqrt{\sigma_m}}{\pi^{1/4}}\right)^{2^m-1}\right\vert^2\notag\\
&=\frac1N\int_{-\eta}^{\eta}\di{\bf s}\left(\frac{\sigma_m}{\sqrt\pi}\right)^{2^m-1}\sum_{i=0}^{2^m}\binom{2^m}{i}^2\notag\\
&=\frac1N(2\eta)^{2^m-1}\left(\frac{\sigma_m}{\sqrt\pi}\right)^{2^m-1}\binom{2^{m+1}}{2^m}\notag\\
&=\frac1{2^{2^m}}\binom{2^{m+1}}{2^m}\left(\frac{2\eta\sigma_m}{\sqrt\pi}\right)^{2^m-1}.
\end{align}
In Table~\ref{tabGKPBinom} we show the expected success probabilities for different values of $m$. To give an idea we study the limit of large $m$ where we have $\binom{2^{m+1}}{2^m}\sim4^{2^m}/(\sqrt\pi\sqrt{2^m})$. It means that the success probability behaves asymptotically as:
\beq
p_m(0)\sim\frac1{2\sqrt{2^m}}\left(\frac4{\sqrt\pi}\right)^{2^m}(\eta\sigma_m)^{2^m-1}.
\eeq
Recall that this expression is valid under the constraint that $\eta\cdot\sigma_m\ll1$. So in particular $\eta\sigma_m\cdot4/\sqrt\pi<1$, which ensures that the probability remains well-defined. Also note that the notion of scaling is absent from these considerations. Both $\eta$ and $\sigma_m$ are fixed parameters here.  

\begin{figure}[h]
\begin{tabular}{|c|c|c|c|}
\hline
$m$  & Squeezing  & Overlap with & Success \\
 & equivalent & Gaussian GKP & probability\\
\hline
1  & 5 dB & 0.9976 & $1.7\eta\sigma$\\
2  & 8 dB & 0.9986 & $6.3(\eta\sigma)^3$\\
3  & 11 dB & 0.9997 & $1.2\cdot10^2(\eta\sigma)^7$\\
4  & 14 dB & 0.9999 & $5.6\cdot10^4(\eta\sigma)^{15}$\\
\hline
\end{tabular}\caption{Comparison between binomial GKP states and their closest Gaussian GKP counterpart and success probability of the protocol to generate them, according to Eq.\eqref{eqSucProba}. The overlap is determined using Eq.\eqref{eq:m-vs-sigma}. Recall that the number of peaks scales as $2^m+1$ and that the results derived here assume $\eta\sigma<1$.\label{tabGKPBinom}}
\end{figure}
\subsubsection{Impact on the output state}

We now focus on the quality of the output binomial GKP state after measuring the $2^m-1$ ancillary modes. We still consider that the resolution obeys the two conditions mentioned above, namely $\eta\cdot2^{m-1}\ll1$ and $\eta\cdot\sigma_m\ll1$. We can show that in this case the state is actually exactly the correct pure binomial GKP state. We are going to derive this explicitly for the two-mode scenario depicted in Fig.~\ref{figGlancyCat}, which corresponds to the binomial GKP state made of three peaks.

We use the mapping in Eq.\eqref{eqBS} for two identical squeezed cat states at the input $\ket{\sqrt2\sqrt\pi,\sigma}+\ket{-\sqrt2\sqrt\pi,\sigma}$, where $\sigma<1$. We obtain a two-mode entangled state that reads -- with the convention that the first/second ket denotes the upper/lower mode:
\begin{align}
\ket{\psi_1^{\rm bm}} & =\frac1{\sqrt N}\left[\ket0\ket{-2\sqrt\pi ,\sigma}+\left(\ket{2\sqrt\pi ,\sigma}+\ket{-2\sqrt\pi ,\sigma}\right)\ket0  \right. \notag \\ 
 & \left.+\ket0\ket{2\sqrt\pi ,\sigma}\right],
\end{align}
where $N$ is the normalization constant.
The upper mode is measured in the $\hat p$ quadrature and the outcome $0$ is recorded. If we suppose that the resolution satisfies $\eta\cdot2\sqrt\pi =o(1)$ and $\eta\cdot\sigma=o(1)$, then Eq.\eqref{eqProj} still holds. With the notations of this section we have for all $s\in[-\eta,\eta]$ and all $\beta=0,\pm2\sqrt\pi $:
\beq\label{eqProjSq}
_p\langle s\ket{\beta,\sigma}\approx\frac{\sqrt\sigma}{\pi^{1/4}}.
\eeq
Recall that we assume that the displaced squeezed vacuum states are orthogonal to each other. Then the reduced density matrix for the mode being measured $\rho_r$ after tracing over the unmeasured mode reads:
\begin{align}
\rho_r&={\rm Tr_{\rm low}}\left[\ket{\psi_1^{\rm bm}}\bra{\psi_1^{\rm bm}}\right]\notag\\
&=\frac14\left[2\ket{0,\sigma}\bra{0,\sigma} \right.\notag\\
&+\left. \left(\ket{2\sqrt\pi,\sigma}+\ket{-2\sqrt\pi,\sigma}\right)\left(\bra{2\sqrt\pi,\sigma}+\bra{-2\sqrt\pi,\sigma}\right)\right],
\end{align}
so that the success probability yields, under the assumptions leading to Equation~\eqref{eqProjSq}:
\begin{align}
p_1(0)&=\int_{-\eta}^{\eta}\di s\bra{ s}\rho_r\ket{ s}\notag\\
&=\frac14\int_{-\eta}^{\eta}\di s\left[2\left\vert_p\langle s\ket{0,\sigma}\right\vert^2+\left\vert_p\bra s\left(\ket{2\sqrt\pi,\sigma}+\ket{-2\sqrt\pi,\sigma}\right)\right\vert^2\right]\notag\\
&=\frac14\left[2\cdot2\eta\frac{\sigma}{\sqrt\pi}+4\cdot2\eta\frac{\sigma}{\sqrt\pi}\right]\notag\\
&=\frac{3\eta\sigma}{\sqrt\pi}.
\end{align}
We can easily check that this is consistent with the general formula derived in Eq.\eqref{eqSucProba}. We may now compute the output state after obtaining outcome 0 at the measurement. It reads:
\begin{align}
\rho_1=\frac{{\rm Tr_{\rm up}}\left[P_1(0)\ket{\psi_1^{\rm bm}}\bra{\psi_1^{\rm bm}}\right]}{p_1(0)},
\end{align}
where $P_1(0)=\int_{-\eta}^{\eta}\di s\ket s_p\bra s$ is the projector associated with outcome 0 and the upper mode. Then we have:
\begin{align}
\rho_1 &=\frac{1}{p_1(0)}  \left(\int_{-\eta}^{\eta}\di s_p\langle s\ket{\psi_1^{\rm bm}}\bra{\psi_1^{\rm bm}}s\rangle_p\right)\notag\\
& =\frac{1}{6}  \left(\ket{2\sqrt\pi,\sigma}+ 2  \ket{0,\sigma}+\ket{-2\sqrt\pi ,\sigma} \right) \notag \\
& \times \left(\bra{2\sqrt\pi,\sigma}+ 2  \bra{0,\sigma}+\bra{-2\sqrt\pi ,\sigma}\right),
\end{align}
which is indeed a pure state corresponding to the definition of the first binomial GKP state. This reasoning can be extended by induction to show that binomial GKP states are pure at all orders for a good enough resolution. 

Now that we have studied and characterized the binomial GKP states that can be generated using Schr\"odinger cat states we will describe a protocol that approximately generates the latter using vacuum states.

\subsection{Cat states from coherent states}\label{secCatKerr}

The protocol for binomial GKP state generation that we have outlined above is based on the use of cat states of the form
\beq
\label{eq:cat}
  \frac{1}{\sqrt{N}} (\ket{\alpha} + \ket{- \alpha}),
\eeq
In the following we assume that all coherent states amplitudes are large enough so that the states can be considered as orthogonal. In particular it means that $N=2$ in Eq.\eqref{eq:cat}.

In this section we  detail how these states can be generated probabilistically and approximately given a set of elementary gates and homodyne detection.
We use the probabilistic protocol of Ref.~\cite{Vitali1997}.
A $\pi$-strength cross-Kerr interaction acting on two coherent states of amplitudes $ \ket{\alpha_0}$ and $\ket{\beta_0}$ yields
\beq
\label{eq:crossKerr}
 e^{i \pi \hat n_1 \hat n_2}  \ket{\alpha_0} \ket{\beta_0}  = \frac{1}{2} (\ket{\alpha_0} + \ket{- \alpha_0}) \ket{\beta_0} +  \frac{1}{2} (\ket{\alpha_0} - \ket{- \alpha_0}) \ket{-\beta_0} .
\eeq
We see from Eq.(\ref{eq:crossKerr}) that if by measuring the second mode one can infer an amplitude of $\beta_0$, then the first mode is projected onto the cat state we are interested in given by Eq.(\ref{eq:cat}), with coherent components of amplitude $\alpha_0$.
Since the probability density for homodyne detection on the second mode consists of two Gaussians at $\pm \beta_0$ with a very small overlap $\bra{-\beta_0} \beta_0 \rangle$, this measurement, even with a poor resolution, will be able to discriminate between $\ket{\pm \beta_0} $.

We need hence a viable decomposition of the cross-Kerr operator  $e^{i \pi \hat n_1 \hat n_2}$ appearing in Eq.(\ref{eq:crossKerr}) in terms of elementary gates from the universal set. 
Crucially, the error with which this gate has to be implemented must lie below a fixed threshold, which is related to the desired fidelity on the resulting GKP states. 
As we will see, this threshold imposes constraints on how the decomposition should be performed. 

Several strategies for gate decomposition are possible. In Refs.~\cite{Suzuki1990, Yoshida1990} an alternate technique to {\it Trotterization} is presented, referred to as {\it splitting}, or fractal decomposition.
This method is particularly suitable when the parameter characteristic of the interaction is small, otherwise very high-order decompositions are needed (see condition (44) in Ref.~\cite{Suzuki1990}). 
Inspired by Ref.~\cite{Sefi2011}, we use a hybrid ad-hoc strategy that combines these two approaches, preceded by concatenation.  Our decomposition is  thereby structured in the following nested steps, that progressively reduce the strength of the interaction applied: (1) Concatenation, (2) Splitting, and (3) Rescaling. In the following, we describe the decomposition in detail. We use a different order of presentation of the steps listed above, in order to make the presentation clearer.
\subsubsection{Decomposition of the cross-Kerr interaction}


\paragraph{Splitting.}

We aim at implementing the cross-Kerr evolution $e^{i \pi  \hat n_1 \hat n_2}$ to a total final precision $y$. In order to do so, the operator $\hat n_1 \hat n_2$ defining the cross-Kerr interaction must be split in operators belonging to the set of elementary gates. First note that the cross-Kerr operator is given by
\begin{align}
\label{eq:cross-kerr}
 \hat n_1 \hat n_2   &= \frac{1}{2} (\hat q^2 + \hat p^2 -1)_1 \otimes \frac{1}{2} (\hat q^2 + \hat p^2 -1)_2 \\
&= \frac{1}{4} \lt \hat q^2_1 \hat q^2_2 +  \hat q^2_1 \hat p^2_2 +  \hat p^2_1 \hat q^2_2 +  \hat p^2_1 \hat p^2_2 \rt -  \frac{1}{4} (\hat q^2_1 +  \hat p^2_1)-  \frac{1}{4} (\hat q^2_2 +  \hat p^2_2) \notag \\
&\equiv  \hat O_1 + \hat O_2 + \hat O_3 + \hat O_4 -  \frac{1}{4} (\hat q^2_1 +  \hat p^2_1)-  \frac{1}{4} (\hat q^2_2 +  \hat p^2_2) \notag
\end{align}
%
where we have defined
$\hat O_1 = \hat q^2_1 \hat q^2_2 /4$, $\hat O_2 = \hat q^2_1 \hat p^2_2  /4$, $\hat O_3 =  \hat p^2_1 \hat q^2_2  /4$, $\hat O_4 =  \hat p^2_1 \hat p^2_2 /4$. Note the operator $\hat O_1+\hat O_2+\hat O_3+\hat O_4$ commutes with the product of the Fourier transforms acting on modes 1 and 2. So the cross-Kerr evolution of amplitude $\pi$ reads 
\begin{equation}
e^{i \pi  \hat n_1 \hat n_2  } = e^{i \pi ( \hat O_1 + \hat O_2 + \hat O_3 + \hat O_4)} \hat F^{\dagger}_1 \hat F^{\dagger}_2. 
\end{equation}
Therefore, we can focus on the decomposition of the operator $e^{i \pi ( \hat O_1 + \hat O_2 + \hat O_3 + \hat O_4)}$. When the parameter $\tau$ characteristic of the interaction is smaller than one, 
applying twice the second-order splitting $e^{i \tau (A + B)} = e^{\frac{i \tau}{2} A} e^{i \tau B} e^{\frac{i \tau}{2} A}  + O(\tau^3)$~\cite{Suzuki1990} gives
\begin{align}
\label{eq:second-order-splitting}
 &e^{i \tau ( \hat O_1 + \hat O_2 + \hat O_3  + \hat O_4)} = \\
 & e^{i \frac{\tau}{4} \hat O_1 } e^{i \frac{\tau}{2}  \hat O_2 }   e^{i \frac{\tau}{4} \hat O_1 } e^{i \frac{\tau}{2} \hat O_3 }  e^{i 
 \tau \hat O_4 } e^{i \frac{\tau}{2} \hat O_3 } e^{i \frac{\tau}{4} \hat O_1 } e^{i \frac{\tau}{2}  \hat O_2 }   e^{i \frac{\tau}{4} \hat O_1 }  \notag \\
   &+ \tau^3 f(\hat O_1, \hat O_2, \hat O_3, \hat O_4). \notag
 \end{align}

\paragraph{Concatenation.}

Overall however, we need a strength $e^{i \pi \hat n_1 \hat n_2}$ for the cross-Kerr interaction as appearing in Eq.(\ref{eq:crossKerr}).
In order to achieve this, we observe that
\beq
\label{eq:amplification}
 e^{i \pi \hat n_1 \hat n_2} = \left( e^{i \frac{\pi}{p} \hat n_1 \hat n_2} \rt^p.
\eeq
Identifying $\tau =\pi/p$, from Eq.(\ref{eq:second-order-splitting}) we see that the precision with which each step  $\lt e^{i \frac{\pi}{p} \hat n_1 \hat n_2} \rt$ is implemented is $\tau^3 = (\pi/p)^3$.  %
From Eq.(\ref{eq:amplification}), the desired gate $e^{i \pi \hat n_1 \hat n_2}$ is then implemented up to precision $p \tau^3 = \pi^3/p^2$.  This must equal the total desired precision $y$. Therefore, this allows us to fix how the number of needed iterations  $p$ depends on the precision $y$:
\beq
p =\lt  \frac{\pi^3}{y} \rt^{\frac{1}{2}},
\eeq
yielding
\beq
\label{tau-vs-y}
\tau = \frac{\pi}{p} =\lt \frac{y}{\pi}  \rt^{\frac{1}{2}}.
\eeq

\paragraph{Rescaling.}

We now have to decompose the factors appearing in Eq.(\ref{eq:second-order-splitting}).
From Eq.(5) of Ref.~\cite{Sefi2011} we have~\footnote{Eq.(5) of Ref.~\cite{Sefi2011} is derived with different conventions for the commutator $\left[ \hat q , \hat p \right]$ from ours. Therefore, the numerical pre-factor should change. However, due to a missing factor $4$ in the numerator of the lhs in Eq.(5) of Ref.~\cite{Sefi2011}, that equation is correct within our notations and we can use it as it is.}
\beq
\label{eq:p1p2}
\hat p^2_1 \hat p^2_2 = - \frac{1}{9} \left[ \hat p^3_2 , \left[ \hat p^3_1 , \hat q_1 \hat q_2 \right] \right]. 
\eeq
The evolution stemming from the operator $\hat q_1 \hat q_2$ belongs to our set of gates, while those corresponding to $p^3_2$ and $p^3_1$ can be obtained by Fourier-transforming the cubic phase gate also in the set, namely $\hat p^3 = \hat F \hat q^3 \hat F^{\dagger}$.
The other terms of Eq.(\ref{eq:cross-kerr}) can be obtained by Fourier transforming on one or two modes, e.g. 
\beq
e^{\frac{i \tau \hat q^2_1 \hat q^2_2}{4}} =\hat  F_1^{\dagger}  \hat F_2^{\dagger} e^{ \frac{i \tau \hat p^2_1 \hat p^2_2}{4}} \hat F_1 \hat  F_2.
\eeq
Therefore, it is only necessary to further decompose the operator in Eq.(\ref{eq:p1p2}), namely the evolution $e^{ \frac{i \tau \hat p^2_1 \hat p^2_2}{4}} = e^{- \frac{i \tau \left[ \hat p^3_2 , \left[ \hat p^3_1 , \hat q_1 \hat q_2 \right] \right]}{36}}$.

Rescaling can be carried out as done in the supplementary material of Ref.~\cite{Sefi2011}. In Appendix \ref{app:nested} we derive the rescaling equations relevant to our purposes, in analogy to the derivation presented in Ref.~\cite{Sefi2011}. We obtain
\begin{align}
\label{eq:nested}
e^{i \tau  \left[ \hat B, \left[ \hat C,\hat A \right] \right]} &= \lt e^{i \hat B \frac{\tau^{1/3}}{k}} \lt e^{i \hat A \frac{\tau^{1/3}}{k l }} e^{i \hat C \frac{\tau^{1/3}}{k l }} e^{-i \hat A \frac{\tau^{1/3}}{k l }} e^{-i \hat C\frac{\tau^{1/3}}{k l }}    \rt^{l^2} \right. \notag \\
& \left. e^{-i \hat B \frac{\tau^{1/3}}{k}}  \lt e^{-i \hat A \frac{\tau^{1/3}}{k l }} e^{-i \hat C \frac{\tau^{1/3}}{k l }} e^{i \hat A \frac{\tau^{1/3}}{k l }} e^{i \hat C\frac{\tau^{1/3}}{k l }}    \rt^{l^2}  \rt^{k^3} \notag \\
& + O \lt \frac{\tau^{4/3}}{k} \rt + O \lt \frac{\tau}{l} \rt,
\end{align}
which we can now use (upon a further rescaling $\tau \rightarrow - \tau /36$) with the following identification of the operators in Eq.(\ref{eq:p1p2}): $\hat A=\hat q_1 \hat q_2$ and $\hat B= \hat p_2^3$ and $C = \hat p_1^3$. Since in Eq.(\ref{eq:second-order-splitting}) there appear $9$ operators that require a decomposition of the kind in Eq.(\ref{eq:nested}), in order to provide a more conservative estimate we require that each factor is implemented up to a precision $\tau^3/9$. Therefore we must require the identifications 
$\tau^{4/3}/ (36^{4/3} k) \sim  \tau^3 / 9$ and $\tau/ (36 l) \sim \tau^3 /9$ ,  which implies using Eq.(\ref{tau-vs-y})
\begin{align}
\label{eq:l-and-m}
k &\sim 9^{-\frac{1}{3}} 4^{-\frac{4}{3}} \tau^{-\frac{5}{3}}  = 9^{-\frac{1}{3}}4^{-\frac{4}{3}} \lt \frac{y}{\pi} \rt^{-\frac{5}{6}} \notag \\
l & \sim \tau^{-2} /4 = \lt \frac{y}{\pi} \rt ^{-1} /4.
\end{align}

We stress again that the final precision $y$ on the cat state generation affects the quality of the binomial GKP states that are generated with this protocol. More precisely, since a binomial GKP state of order $m$ requires $2^m$ cat states, the error $y$ is amplified accordingly: for a given $m$, the distance between the approximate GKP states generated through this procedure and the binomial GKP states is upper bounded by $2^my$. We indicate the approximate binomial GKP states by $\ket{\tilde{0}_m}$. A summary of the fidelities between different GKP states is provided in Table \ref{table-GKPs}.

To summarize, we have provided a decomposition of the cross-Kerr interaction $e^{i \pi \hat n_1 \hat n_2}$ in gates of the form $e^{i b \hat q_1  \hat  q_2}$, $e^{i c \hat q^3}$, $e^{i g (\hat q  \hat p + \hat p  \hat q)}$, $\hat F$.
In Appendix \ref{estimate-number-gates} we provide an estimate of the scaling of the required number of gates in order to achieve a precision of $y$, which results in $N \propto y^{-3}$. This results in particular in a roughly a thousand elementary gates for a required precision of $0.1$ in the cross-Kerr gate implementation.  

What we further have to do is to provide a decomposition of the squeezing gate $e^{i g (\hat q  \hat p + \hat p  \hat q)}$ in terms of the elementary gates of Eqs.(\ref{gates}) and (\ref{gates2}). Also, in the procedure outlined above for the GKP synthesis from cat states a beamsplitter appears. Therefore we must also decompose this transformation into elementary gates. This is the goal of the next subsection.

\subsubsection{Decomposition of beamsplitters and squeezers}

In order to cast the gates used for the cross-Kerr evolution in the form admitted by the set $A_1$ supplemented by the Fourier transform, we have to provide an expression of the squeezing operator and of the beamsplitter in terms of $\hat q$-diagonal gates and Fourier transforms. In contrast to the decomposition of the previous subsection, these can be provided exactly. We provide the detailed calculation in Appendix \ref{app:decomposition-gates-exact}. The result is the following: the squeezing operator is decomposed onto
\beq
\label{eq:decomposition-squeezing}
e^{-i \frac{\ln{s}}{2} (\hat q \hat p + \hat p \hat q )} = \hat F e^{\frac{i s \hat q^2}{2}} \hat F e^{ \frac{i  \hat q^2}{2 s}} \hat F e^{ \frac{i s \hat q^2}{2}},
\eeq
while the beam splitter operator decomposes as
\beq
\label{eq:decomposition-bs}
 \hat F  \hat O( q)  \hat F  \hat O( q)  \hat F   \hat O( q) 
\eeq
with $ \hat O(q) = e^{i \frac{1}{2 \sqrt{2}} (\hat q_1^2 - \hat q_2^2 + \hat q_1\hat q_2 ) }$. Note that in turn $O(\hat q)$ can be trivially decomposed into shear and $\hat C_Z$ gates.
Now all the gates necessary to implement an approximate GKP state have been decomposed onto gates of the elementary set Eqs.(\ref{gates}) and (\ref{gates2}).

\begin{figure}[h]
\hspace{-0.5cm}
\begin{tabular}{c c c c c}
\hline
Approximate GKP & $\rightarrow$ & Binomial GKP & $\rightarrow$ & Gaussian GKP \\
$\ket{\tilde{0}_m}$  &     $2^m y$             & $\ket{0_m}$ & $\zeta_m$ & $\ket{0_G}$ \\
\hline
\end{tabular}\caption{Summary of the definitions of the different fidelities between GKP states addressed in this paper. The overall distance between $\ket{\tilde{0}_m}$ and $\ket{0_G}$ is $\varepsilon_{\rm m}=\zeta_m+2^my$. Also note that the respective encoded data qubits share the same trace distances, due to the fact that encoding consists of a unitary operation. \label{table-GKPs}}
\end{figure}

\section{Noise in CV and fault tolerance}\label{secFTUQC}

Fault tolerance in CV is an issue that must be addressed specifically. In this section we show that the universal model defined in \ref{definition-of the modelsA} can be made fault-tolerant.

\begin{figure}[h]
\includegraphics[width=0.9\columnwidth]{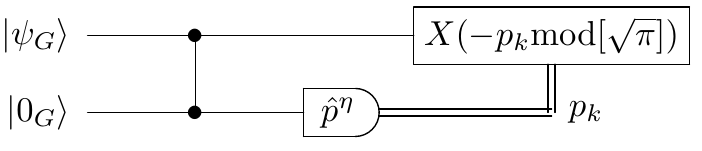}

\caption{\label{figECGadget}Procedure to correct for errors in the $\hat q$ quadrature. $\ket\psi$ is the data qubit and $\ket{0_G}$ is a Gaussian GKP state. After measurement on the second mode the result $p_k$ is used to shift the first mode back.}
\end{figure}

In CV, the natural basis for quantum channels consists of all possible displacement operators. Formally, a general noise model $\mathcal E$ on an arbitrary input state $\hat\rho$ can be expanded in terms of shifts acting on $\hat\rho$, according to the following expression~\cite{Gottesman2001}:
\beq
\label{eq:modelization-noise}
\mathcal E(\hat\rho)=\int\di u\di v\di u'\di v'C(u,v,u',v')e^{-i u \hat{p}} e^{-i v \hat{q}}\hat\rho e^{iv'\hat q}e^{iu'\hat p},
\eeq
where $\mathcal E$ is completely positive and trace preserving. Therefore an error correcting procedure in CV applies to arbitrary noise models if it allows one to correct for arbitrary displacement errors (similar to qubits where an arbitrary error can be corrected based on Pauli-error correction), which intuitively is possible when the support of the error distribution $C$ is concentrated on sufficiently small values. This feature is enabled by the GKP encoding~\cite{Gottesman2001} using an additional source of ancillary GKP $\ket{0_G}$ states, where as we have already noted the notation $\ket{0_G}$ indicates that the ancillary GKP states employed are the Gaussian ones, i.e. enveloped by a Gaussian and correspondingly associated with a finite squeezing degree. The specific circuit for error correction is shown in Fig.~\ref{figECGadget}.

The noise reduction (i.e. error correction in the CV sense) works if the measurement on the ancillary mode yields the correct $\sqrt\pi$-long interval on the real axis. The displacement error acting on the data qubit in one of the quadratures is then replaced by the independent error coming from the fresh ancillary GKP state. Repeating the procedure after a Fourier transform ensures that the displacement errors in both quadratures are replaced. To summarize, this probabilistic procedure ensures that the CV noise remains controlled and determined by the noise linked to the supply of GKP states. The success probability itself is determined by how likely the noise is to displace the state by more than $\sqrt\pi/2$.

An important step was taken in~\cite{Menicucci2014}, where the authors combined this noise reduction procedure with an additional layer of qubit error-correcting code at the logical encoding level. They showed that if the procedure fails -- that is the measured displacement error was larger than $\sqrt\pi/2$ -- the controlled-displacement actually corresponds to a bit flip at the GKP encoding level. Then one can use a standard qubit error-correcting code to address this bit flip error. This ensures that the failure probability of the procedure, as soon as it is lower than the threshold associated with the qubit error-correcting code, can be made arbitrarily small, thus enabling fault tolerance for a CV hardware.  
Note that the squeezing level required in order to fulfill the threshold condition according to the procedure described above can be lowered by utilizing a more refined error correction scheme~\cite{Fukui2018}. The latter exploits a priori information stemming from the Gaussian distribution in each GKP peak together with the analog information associated with the syndrome detections, rather than only using binary information corresponding to the measured displacement being `inside' or `outside' the above mentioned $\sqrt{\pi}$-long interval. This method also allows for the reduction of the number of the concatenation steps in a full error-correcting procedure~\cite{Fukui2018b}. For the purpose of keeping our discussion less technical, however, we will refer to the original error correction scheme of Ref.~\cite{Menicucci2014} in order to derive bounds on the squeezing level as well as other parameters for our model.

In practice then, the success probability is determined by how likely the noise is to displace the state by more than $\sqrt\pi/2$. There may be numerous origins for this CV noise: usually not only the ``data qubit'' will be noisy, but the ancillary one and the measurement (e.g. finite resolution) will also. Crucially the characteristics of the noise do not really matter. The only relevant questions are: what is the probability that all these errors put together yield a displacement larger than $\sqrt\pi/2$? Is it possible to make this probability lower than the threshold of some qubit error-correcting code $\varepsilon_{\rm th}$? Mathematically, we aim to analyse and upper bound the following probability:
\beq\label{eqProbaEC}
{\rm Pr}\left(\vert p\vert>\frac{\sqrt\pi}2\right),
\eeq
where $p$ is the displacement error measured in the protocol shown in Fig.~\ref{figECGadget}. However this probability may turn out to be too difficult to compute. We show in the following how to reduce this problem to a much simpler one only involving Gaussian functions.

We are going to assume GKP encoding as explained above. The data qubit is, however, encoded in the approximate GKP states basis obtained with the procedure outlined in Sec.~\ref{secGKP}. We denote the corresponding density matrix with $\tilde{\rho}_m$. 
When noise enters the picture the output state $\rho$ can be a mixed state. We characterize the noise model by a parameter $\varepsilon$ which upper bounds the distance:
\beq
d(\rho,\tilde{\rho}_m)\equiv\frac12\norm{\rho-\tilde{\rho}_m}<\varepsilon
\eeq  
where $\norm{\cdot}$ indicates the standard trace norm.
Similarly we can upper bound the distance between the approximate GKP states and the standard Gaussian GKP states $\ket{0/1_G}$. Recall that it is determined by two contributions: $\zeta_m$ coming from the imperfect fidelity between the binomial GKP states and the Gaussian GKP states -- see Table~\ref{tabGKPBinom}, Table~\ref{table-GKPs} and Eq.\eqref{eqOverlap}; and $2^my$ because of the approximate procedure used to generate Schr\"odinger cat states and thus approximate GKP states, stemming from finite precision gates -- see the discussion following Eq.\eqref{eq:l-and-m}. Sticking with the density matrix formalism for consistency, we denote as $\rho_G$ the closest Gaussian GKP-encoded state to the corresponding approximate GKP-encoded state $\tilde{\rho}_m$. :
\beq\label{eqepsm}
d(\tilde{\rho}_m,\rho_{G})<\varepsilon_m,
\eeq
where $\varepsilon_m=\zeta_m+2^my$.

Suppose now that one wishes to reduce the noise of a state $\rho$ using the procedure described in Fig.~\ref{figECGadget}. There, the input state is $\rho_G  \ket{0_G}\bra{0_G}$, where the tensor product is implicit. In our scheme, the input state is replaced by $\rho \ket{\tilde{0}_m} \bra{\tilde{0}_m}$, where $\ket{\tilde{0}_m}$ is the logical 0 state encoded in the approximate GKP basis.
We focus first on the trace distance for the joint state after the $C_Z$ gate shown in Fig.~\ref{figECGadget}. For a perfect, unitary $C_Z$ gate and using properties of the trace norm we have:
\begin{align}\label{eqdistance}
d(C_Z\rho\ket{\tilde{0}_m}&\bra{\tilde{0}_m} C_Z^{\dagger},C_Z\rho_G\ket{0_G}\bra{0_G}C_Z^{\dagger})\notag\\
&=d(\rho\ket{\tilde{0}_m}\bra{\tilde{0}_m},\rho_G\ket{0_G}\bra{0_G})\notag\\
&\leq d(\rho\ket{\tilde{0}_m}\bra{\tilde{0}_m},\tilde{\rho}_m\ket{\tilde{0}_m}\bra{\tilde{0}_m})\notag\\
&\quad+d(\tilde{\rho}_m\ket{\tilde{0}_m}\bra{\tilde{0}_m},\rho_G\ket{\tilde{0}_m}\bra{\tilde{0}_m})\notag\\
&\quad+d(\rho_G\ket{\tilde{0}_m}\bra{\tilde{0}_m},\rho_G\ket{0_G}\bra{0_G})\notag\\
&<\varepsilon+ \varepsilon_m + \varepsilon_m.
\end{align}
In other words, the statistical distance for the noise reduction protocol using approximate GKP states as ancillary qubits compared to using Gaussian GKP states is upper bounded by $\varepsilon+2\varepsilon_m$. Hence the possible deviations of the measurement results are also upper bounded by $\varepsilon+2\varepsilon_m$. 

We now define $p$ ($p_G$) as the measured displacements in the (Gaussian) protocol. Eq.\eqref{eqdistance} implies that $\vert p-p_G\vert<\varepsilon+2\varepsilon_m$. Let us consider now the failure probability that we wish to upper bound, that is
\begin{align}
\label{eq:failure-prob}
P_{\text{fail}} = {\rm Pr}\left(\vert p\vert>\sqrt\pi\right)&={\rm Pr}\left(\vert p-p_G+p_G\vert>\frac{\sqrt\pi}2\right)  \\
&\leq{\rm Pr}\left(\vert p_G\vert+\vert p-p_G\vert>\frac{\sqrt\pi}2\right)\notag\\
&\leq{\rm Pr}\left(\vert p_G\vert>\frac{\sqrt\pi}2-\vert p-p_G\vert\right)\notag\\
&\leq{\rm Pr}\left(\vert p_G\vert>\frac{\sqrt\pi}2-(\varepsilon+2\varepsilon_m)\right) \notag.
\end{align}
In other words, if we aim to upper bound the probability that our error correction protocol fails -- and show that it can be made lower than some threshold probability $\varepsilon_{\rm th}$ -- we simply have to consider the Gaussian protocol using $\sqrt\pi/2-(\varepsilon+2\varepsilon_m)$ as a bound rather than $\sqrt\pi/2$. 

Recall finally that the protocol actually needs to be repeated after a Fourier transform to cover for the noise in both quadratures. The protocol is successful if both rounds are, i.e. the total failure probability can be expressed as
\beq
\label{eq:failure-prob2}
1 - P_{\text{succ}_1} P_{\text{succ}_2} = 1 - (1-P_{\text{fail}_1})(1- P_{\text{fail}_2}),
\eeq
where $1$ and $2$ refer to error correction rounds on the two respective quadratures. 
For real $\sigma$ and $\delta$ we denote $\chi(\sigma,\delta)={\rm Pr}\left(\vert p_G(\sigma)\vert>\sqrt\pi/2-\delta\right)$, where we stress that $p_G$ actually depends on the squeezing $\sigma$ of the Gaussian GKP states. A given noise model corresponds to non-vanishing values for the distances $\varepsilon$ and $\varepsilon_m$. Then, combining Eq.(\ref{eq:failure-prob}) and (\ref{eq:failure-prob2}), the fault tolerance condition amounts to finding a squeezing parameter $\sigma_{\rm th}$ such that the following bound on the total failure probability holds:
\beq\label{eqThreshold}
1-\left(1-\chi(\sigma_{\rm th},\varepsilon_q+2\varepsilon_m)\right)\left(1-\chi(\sigma_{\rm th},\varepsilon_p+2\varepsilon_m)\right)<\varepsilon_{\rm th}.
\eeq
Note that in principle the Fourier transform in between the two rounds of noise reduction may disturb the system and thus yield a different upper bound for the trace distance, hence the notation $\varepsilon_{1,2}$ -- e.g. the Measurement Based implementation of the Fourier transform (see e.g. the Supplementary Information of Ref.~\cite{Douce2017} for a detailed discussion).

To summarize we have shown how to reduce an arbitrary noise model to simply considering Gaussian GKP states with slightly modified thresholds. This allows one to easily compute the relevant probabilities since they now only consist of integrating tails of Gaussian functions. 

\section{Consequences for the CV gates}
\label{sec-consequences}

In this section we determine the conditions that fault tolerance requirements impose on the gate parameters defined in Sec.~\ref{definition-of the models} Eqs.\eqref{gates} and \eqref{eq:gates-sub2}. These also depend on the precision of the gate decomposition that we derived in the previous sections.
\subsection{Universal computational model}
\label{secUniv}

The results detailed in Section \ref{secFTUQC} pave the way for fault tolerance: whatever noise model one has to deal with, it can be plugged in our computational model to be translated into specific requirements for the gate parameters and still allow for error correction.

Here, we find instructive to discuss the case given in the abstract definition of our computational model introduced in Sec.~\ref{definition-of the modelsA}, where the gates belonging to the model are assumed to be perfectly implemented, i.e. there are no external sources of error. This corresponds to setting $\varepsilon_q$ = $\varepsilon_p$ = 0 and no active error correction is required. 
However, using approximate GKP states for logical information encoding yields an intrinsic error probability. Following the analysis performed in~\cite{Gottesman2001}, this probability can be linked to the integrals of Gaussian functions, i.e. to the error function. For a Gaussian GKP state of symmetric squeezing parameter $\sigma$ it simply reads ${\rm erf}(\sqrt\pi/2\sigma)$, where ${\rm erf}$ denotes the error function. In our case this probability also depends on the $\varepsilon_m$ defined in Eq.\eqref{eqepsm}, as well as on the gate parameters required for universal quantum computing. Namely, we have:
\beq
\label{eq:psucc-m-y}
P_{\rm succ}(m,y)={\rm erf}\left(\frac{\frac{\sqrt\pi}{2}-\varepsilon_m}{\sigma}\right),
\eeq
where $m$ and $\sigma$ are related according to the discussion in Sec.~\ref{secGKP}, and where in particular $\varepsilon_m$ depends on the precision $y$.

\begin{figure}[h]
\includegraphics[width=\columnwidth]{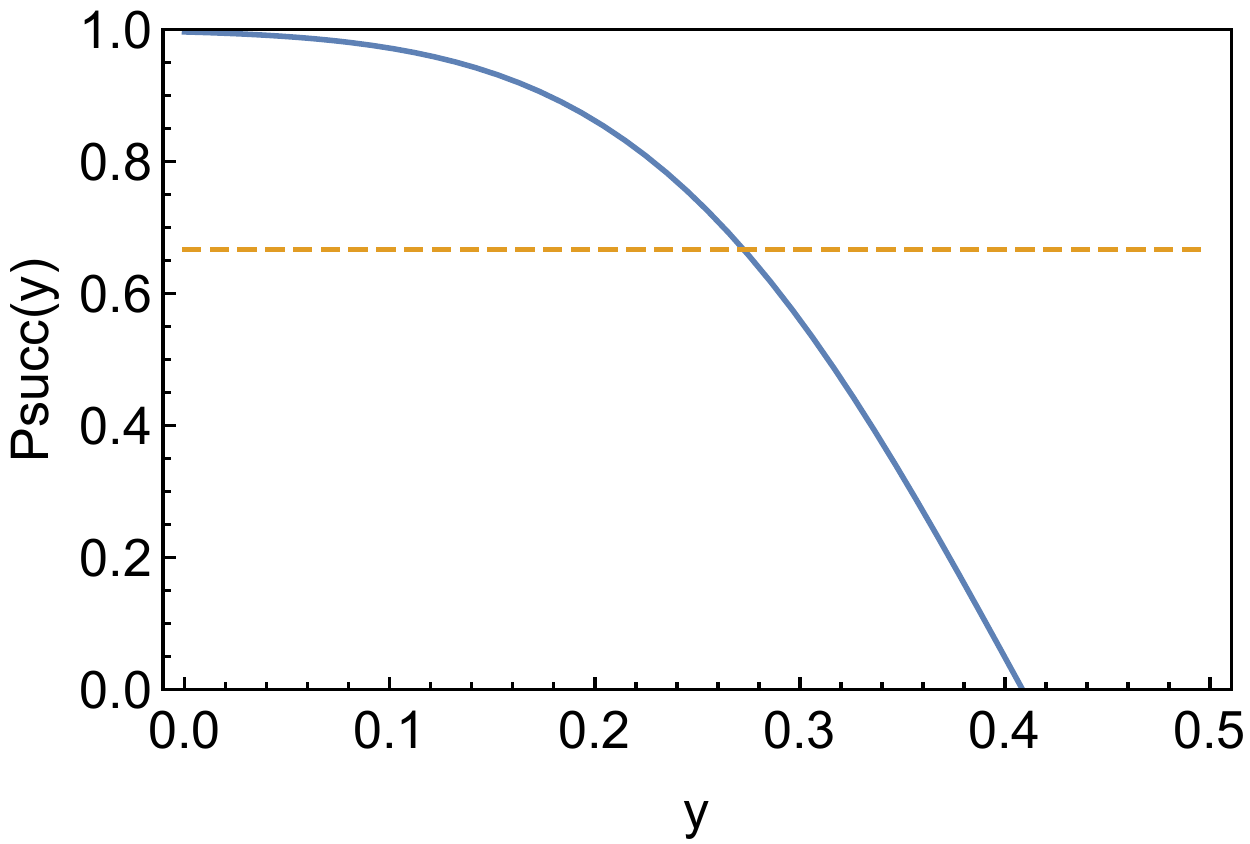}
\caption{Success probability $P_{\rm succ}$ in Eq.(\ref{eq:succ-prob-y}) as a function of $y$ (solid line). As a guide for the eye, the reference level $2/3$ is also plotted (dashed line).\label{figYPlot}}
\end{figure}

We consider the experimentally easiest case where $m=1$, so $\varepsilon_1=\zeta_1+2y$. Given the analysis performed in Sec.~\ref{secGKP}, the corresponding approximate GKP states can be compared to Gaussian GKP states with a squeezing parameter of 5 dB. The associated fidelity shown in Table~\ref{tabGKPBinom} can be used to estimate $\zeta_1$:
\beq
\zeta_1\leq\sqrt{1-0.9976^2}\approx0.069.
\eeq
In order to compute the corresponding gate parameters, we plot in Fig.~\ref{figYPlot} the success probability $P_{\rm succ}$ as a function of the errors $y$ in the implementation of the cross-Kerr interaction discussed in Sec.~\ref{secGKP}. It reads:
\beq
\label{eq:succ-prob-y}
P_{\rm succ}(y)={\rm erf}\left(\frac{\frac{\sqrt\pi}{2}-\zeta_1-2y}{\sigma}\right).
\eeq

To a value of the gate precision of $y = 0.1$ corresponds a success probability of approximately 0.97. Therefore in the following we will use this value to derive the corresponding gate parameters and understand the experimental consequences. 

Based on the requirement of $y=0.1$ and $m = 1$, we can now compute the gate parameters corresponding to the gate set $A_1$ defined in Sec.~\ref{definition-of the modelsA} Eq.\eqref{gates}. As drawn from the previous sections, using in particular Eqs.(\ref{eq:nested}) and (\ref{eq:decomposition-squeezing}) we obtain the numerical values summarized in Table \ref{tab:operations-summary}.
\begin{figure}[h]
\begin{tabular}{|c|c|c|}
\hline
Evolution  & GKP generation step & Parameters \\
\hline
Displacement  & Coherent state initialization  & $d \simeq 5.6$ \\
\hline
Shearing  &   Squeezing of the cat   & $s_1 \simeq 0.73$; $s_2 \simeq 1.1$ \\
\hline
Entanglement & Beamsplitter & $\tilde b_1 \simeq 0.35$ \\
	& Cross-Kerr decomposition & $\tilde b_2\simeq0.011$ \\	
\hline
Cubic & Cross-Kerr decomposition & $c_1 \simeq 0.011$ \\
	&					& $c_2 \simeq 0.086$ \\
\hline
\end{tabular}\caption{Summary of the parameters for the definition of the gates of the universal model Eqs.\eqref{gates} and~\eqref{gates2}, based on a precision of $y = 0.1$ and $m = 1$.  \label{tab:operations-summary}}
\end{figure}
Together with the gates required to implement the beam splitter appearing in Eq.(\ref{eq:decomposition-bs}), this completes the set of gates that defines our model.

\subsection{Subuniversal models}\label{secSub}

In  this section we   provide a proof of classical computational hardness of exact sampling in the worst-case scenario for both subuniversal models introduced in Sec.\ref{definition-of the modelsB}. First we focus on the circuit illustrated in Fig.\ref{figcirc}, and argue that its hardness directly follows as a consequence of the universality of the model of computation that we have addressed in Sec.\ref{secUniv}. Then, we turn to \textsf{CVIQP} (Fig.\ref{figCVIQP}), and refine the proof that was provided in Ref.~\cite{Douce2017} by  dropping the requirement of input GKP states. To achieve this goal, we use similar arguments to those that allowed us to claim fault tolerance in the probabilistic universal model presented above. As we shall see, the required constraints are tighter for the latter model.

\subsubsection{CV random circuit sampling}\label{secSub1}

The first circuit that we address is shown in Fig.\ref{figcirc}, corresponding to picking gates randomly from the sets that define the probabilistic and universal computational model discussed in the previous section~\ref{secUniv}.
For the case of exact sampling, a very standard proof of classical hardness can be provided~\cite{Aaronson2013, Fahri2016, Morimae2014, Douce2017}. In general, if a subuniversal computational model of quantum computation becomes universal when post-selecting on a subset of the outputs -- i.e. contains the class \textsf{PostBQP}, then information-theoretic arguments allow one to conclude that efficient classical simulation of this subuniversal model is impossible. These arguments are based on widely-held conjectures of classical complexity theory, like the so-called Polynomial Hierarchy not collapsing. Hence the main point that we have to show is that post-selection makes the circuit of Fig.\ref{figcirc} as powerful as  \textsf{PostBQP}. 

This statement directly follows from the universality of the model presented in \ref{secUniv}. As we have shown in Sec.\ref{secGKP}, the gates in Eq.(\ref{gates}) allow for the probabilistic generation of approximate GKP states with success probability given by Eq.\eqref{eqSucProba}. Crucially, the fact that the success probability is exponentially low in the number of approximate GKP states that need to be produced does not hinder this result~\cite{AaronsonBlog}. Therefore, any quantum circuit can be run probabilistically using elements drawn only from the family of circuits of Fig.\ref{figcirc}. Then post-selecting on successfully passing these probabilistic events yields the correct computation. Furthermore, one can also post-select on an additional subset of the outputs. It means that universal and post-selected quantum computations -- i.e. \textsf{PostBQP} -- can be performed using circuits of the type described in Fig.\ref{figcirc}.


We also stress that post-selection should be regarded as a mathematical trick for the proof of hardness, and that experimentally probing the model under consideration would just consist in sampling from an architecture belonging to the family depicted in  Fig.\ref{figcirc}, with no actual need for practical post-selection on measurement outcomes, nor for actual generation of GKP states.

\subsubsection{CV Instantaneous Quantum Computing with input squeezed states is hard}\label{secSub2}

The second circuit family we address is sketched in Fig.\ref{figCVIQP}. In order to show that the model \textsf{CVIQP} is hard to sample classically, we   build on the proof presented in~\cite{Douce2017}, which was established for an analogous model, but with Gaussian GKP states available at the input. The squeezing parameter of those GKP states, furthermore, was required to scale logarithmically with the circuit size, namely we had 
\beq
\label{eq:scaling-condition}
\sigma\propto\log n
\eeq
where $\sigma$ in dB characterizes the widths of the Gaussian describing the GKP state wavefunction and $n$ is the number of modes. 
Just like for the model addressed in Sec.\ref{secSub1}, the main structure of reasoning to claim computational hardness was based on showing that the class defined as \textsf{PostCVIQP}, i.e. \textsf{CVIQP} with the additional resource of post-selection, is as powerful as \textsf{PostBQP}. In other words, \textsf{PostCVIQP} is a universal and fault-tolerant model of QC -- supplemented with post-selection. 

In order to obtain universality and fault tolerance in the post-selected model (i.e. to show that it is as powerful as \textsf{PostBQP}), we first show that the condition \eqref{eq:scaling-condition} is actually unnecessary for the hardness statement of Ref.\cite{Douce2017} to hold, and that a constant $\sigma$ is sufficient to prove the hardness of the circuit. Furthermore, note that the GKP state generation can be subsumed in the circuit model definition when this is augmented with post-selection, similarly as we did in \ref{secSub1}. Then, we analyze the bound on the input squeezing imposed by the fault tolerance requirement of the model with post-selection.

The probability of a successful post-selection only makes sense if it is not worse than exponentially low, which in particular is guaranteed by the scaling law for the squeezing parameter derived in \cite{Douce2017}. However, the post-selection happens at a logical, encoding level so it actually corresponds to several physical measurement outcomes recombined according to the error-correcting code. Each of these outcomes is noisy because of the CV nature of the quantum states, which can be understood as having imperfect, non-orthogonal qubits. However, the threshold theorem states that the probability of reaching a wrong conclusion at the encoding level can be made exponentially low since the physical noise can be mapped to a qubit error following the procedure described in the previous section.

More specifically, this mapping is ensured by the use of GKP states, and by associating the measurement outcomes of the homodyne detection in given intervals to either 0 or 1 at the logical level.  For a symmetric Gaussian noise, an upper bound as a function of the squeezing can be found in~\cite{Gottesman2001} with additional approximations. 
Therefore, the scaling condition in Eq.(\ref{eq:scaling-condition}) was actually unnecessary. 

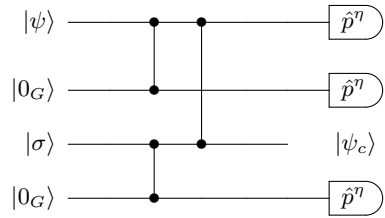
\begin{figure}[h]
$$
\Qcircuit @C=1.7em @R=1.4em {
\lstick{\ket{\psi}}  & \qw & \ctrl{1} & \ctrl{2} & \qw & \qw &  \measureD{\hat p^\eta} \\
\lstick{\ket{0_G}}  & \qw & \control\qw & \qw & \qw & \qw & \measureD{\hat p^\eta} \\
\lstick{\ket{\sigma}}  & \qw & \ctrl{1} & \control\qw & \qw & \qw & \ket{\psi_c} \\
\lstick{\ket{0_G}}  & \qw & \control\qw  & \qw & \qw & \qw &  \measureD{\hat p^\eta} 
}
$$
\caption{\label{figFourier}Circuit implementation of an error-corrected Fourier transform based on the resources available in the \textsf{CVIQP} model, where $\ket{\psi_c}$ denotes the output corrected state. }
\end{figure}

Now, recall that the proof of hardness for \textsf{CVIQP} circuits holds if we can make sure that \textsf{PostCVIQP} circuits are able to realize universal quantum computations. However, since the Fourier transform does not belong to the computational model defined by \textsf{CVIQP} circuits, it can only be implemented in \textsf{PostCVIQP} using a measurement-based approach as shown in Fig.\ref{figFourier}. This approach unfortunately introduces additional noise, which implies that we have to rely on quantum error correction for CV quantum computation as discussed in Section~\ref{secFTUQC}. 

The Fourier transform can be implemented in a single, post-selected, teleportation step. The squeezing parameter $\sigma$ characterizing the squeezed vacuum state is added to the variance of the momentum quadrature of the state being teleported. Based on the analyses developed in Sec.~\ref{secFTUQC} and Ref.~\cite{Menicucci2014} and neglecting the effects of finite resolution, a sufficient condition to achieve fault tolerance can be derived, similar to the one in Eq.\eqref{eqThreshold}, and it reads:
\beq\label{eqThresholdFT}
1-\left(1-\chi(\sqrt2\sigma,2\varepsilon_m)\right)\left(1-\chi(\sqrt5\sigma,2\varepsilon_m)\right)<\varepsilon_{\rm th},
\eeq
for a given qubit error-correcting code of threshold probability $\varepsilon_{\rm th}$.

We chose the lowest value of $m$ compatible with Eq.\eqref{eqThresholdFT}. For a threshold probability $\varepsilon_{\rm th}=10^{-6}$ we find that $m=6$, which corresponds to a squeezing of 19 dB according to Eq.\eqref{eq:m-vs-sigma}, is the minimal amount of squeezing that satisfies Eq.\eqref{eqThresholdFT}. It yields an upper bound on $\varepsilon_6\approx0.05$. Recall that $\varepsilon_6=2^6y+\zeta_6$. Using Eq.\eqref{eqOverlap} we have $\zeta_6\approx3.10^{-3}$ so we get an upper bound for the error rate in the Schr\"odinger cat states generation of
\beq
y\leq10^{-3}.
\eeq
This value constitutes a much stronger requirement than the value of $10^{-1}$ found in Sec.~\ref{secUniv} when discussing the model for universal quantum computing. It stems from the fact that the set of gates in \textsf{CVIQP} circuits does not include the Fourier transform, which must be implemented in a measurement-based fashion entailing an additional error due to finite squeezing in the ancillary squeezed state. Analogous to what we found for the universal model in Table \ref{tab:operations-summary}, this value of precision allows for the derivation of corresponding gate parameters for the definition of the \textsf{CVIQP} model of Eq.(\ref{eq:gates-sub2}). These parameters are shown in Table \ref{tab:operations-summary2}.

\begin{figure}[h]
\hspace{-0.5cm}
\begin{tabular}{|c|c|c|}
\hline
Evolution  & GKP generation step & Parameters \\
\hline
Displacement  & Coherent state initialization  & $\tilde d \simeq 142$ \\
\hline
Shearing &  Squeezing of the cat   & $\tilde s_1 \simeq 0.28$; $\tilde s_2 \simeq 0.89$ \\
\hline
Entanglement & Beamsplitter & $\tilde b_1 \simeq 0.35$ \\
	& Cross-Kerr decomposition & $\tilde b_2\simeq1.6\cdot10^{-6}$ \\	
\hline
Cubic & Cross-Kerr decomposition & $\tilde c_1 \simeq 1.6\cdot10^{-6}$\\
	&		& $\tilde c_2 \simeq 1.3\cdot10^{-3}$ \\
\hline
\end{tabular}\caption{Summary of the parameters for the definition of the gates of the \textsf{CVIQP} model, based on a precision of $y = 10^{-3}$ and $m = 6$.\label{tab:operations-summary2}}
\end{figure}

In conclusion to this section on subuniversal models, we stress that neither for the CV random circuit sampling model discussed in Sec.\ref{secSub1} nor for \textsf{CVIQP} an actual generation of GKP states is needed in practice: the statement on computational hardness holds for sampling from the output probability distributions of the circuits in Fig.\ref{figcirc} and Fig.\ref{figCVIQP}, respectively. In contrast to the universal model, GKP generation and encoding was only used as a conceptual intermediate step for the proof of hardness. 

\section{Conclusions}\label{secCcl}


In summary, we have derived what is to our knowledge the first quantum computational model in CV based only on input vacuum states, and we have shown that a finite set of gates and homodyne detection are sufficient to achieve probabilistic universal QC and fault tolerance. This model can be adapted to yield sampling problems which are hard to sample for classical computers unless the polynomial hierarchy collapses. These consist of a CV random circuit sampling model with the same structure as for universal computations and a model analogous to \textsf{IQP} for qubits based on momentum squeezed vacuum states. 


Regarding fault tolerance, in the universal model the gates are implemented perfectly, and no active correction is required. However, the intrinsic noise inherent to approximate GKP states can be linked to the parameters of the gates defining our model, allowing to determine the value of the gate parameters to use. Consistently with that gate parameters, GKP states generated from two Schr\"odinger cat states can be used, which is experimentally not too demanding. The hardness proof for the CV random circuit model, which is directly based upon the basis of the universal model, relies on similar considerations.

On the other hand, for the \textsf{CVIQP} model, more stringent requirements on fault tolerance are imposed by the fact that the Fourier transform does not belong to the gate set. Correspondingly, stronger constraints on the squeezing parameter and gate precision arise. 

A candidate platform where proof-of-principle experiments of the present models could be addressed is provided by optical systems, where spontaneous parametric downconversion allows for the generation of multi-mode squeezed states of light~\cite{Roslund2013, Roslund_13b}, that can be subsequently manipulated and e.g. entangled in large cluster states~\cite{Chen2014, yokoyama2013optical, Su12}. In these systems, engineering non-linearities necessary for the implementation of cubic gates is challenging. Attempts to probabilistically achieve this aim have been performed by means e.g. of photon subtraction~\cite{Ra2017} and more generally with the use of single photon detectors~\cite{Yukawa2013, Miyata2016}, towards the implementation of a full cubic-phase gate with photon counters~\cite{Gottesman2001, Ghose2007}. More recently, microwave radiation coupled to superconducting qubits has proven useful for the generation of squeezed states~\cite{Wilson2011, Schneider2018}. In this promising set-up, non-linearities allowing for the generation of non-Gaussian states can be provided by the coupling with superconducting Josephson junctions, acting as artificial atoms on the microwave field~\cite{Hofheinz2009}. Finally, opto-mechanical systems are also promising candidates for the generation of multi-mode squeezed and cluster states of radiation ~\cite{Houhou2015}. There, non-Gaussian states could also be deterministically generated~\cite{Brunelli2018}.

We believe our work opens up new perspectives. Concerning the universal model, several aspects can be further optimized, from the gate decomposition~\cite{Suzuki1990} of the cross-Kerr implementation to the error-correcting procedure~\cite{Fujii2016}. 
A further way to drastically reduce the number of gates necessary for the implementation of the cross-Kerr interaction could be to include in the elementary gates set a quartic Hamiltonian $e^{iq^4 s}$~\cite{Sefi2013}.

Furthermore, it would be interesting to study specific implementations of the gates that compose our model. For instance, in Ref.~\cite{Arzani2017} a protocol was given, allowing for the implementation of polynomial approximation of arbitrary-order operations diagonal in the $\hat q$ quadrature (including non-Gaussian operations), with the use of ancillary photon-subtracted squeezed states. Merging the two approaches, upon proper considerations on error-probability and fault tolerance specific to the gate implementation considered, would result in probabilistic fault-tolerant universal QC from photon subtracted squeezed states and Gaussian operations as building blocks.
As for the sampling models considered, it would be desirable to extend the hardness proof to the average case, and to the approximate sampling case.

\section{Acknowledgements}

We acknowledge Nicolas Menicucci and Rafael Alexander for useful discussions, and Seckin Sefi for having noticed a mistake in a precedent version of this manuscript. This work was supported by the ANR COMB project, grant ANR-13-BS04-0014 of the French Agence Nationale de la Recherche. G.F. acknowledges support from the European Union through the Marie Sklodowska-Curie Grant Agreement No. 704192.

\appendix
\section{Nested commutators used for the rescaling step}
\label{app:nested}

The starting point is
\beq
\label{eq:commutator1}
e^{t^2 \lqu \hat B,\hat A \rqu} = e^{i t \hat B}  e^{i t \hat A} e^{-i t \hat B}  e^{-i t \hat A} + O(t^3, \hat A, \hat B).
\eeq
Now using the identity $e^{t^2 \lqu \hat B,\hat A \rqu} = e^{\lt \frac{t}{n} \rt^2 \lqu \hat B,\hat A \rqu n^2}$ and using Eq.(\ref{eq:commutator1}) with $t \rightarrow t/n$ we obtain
\begin{align}
\label{eq:commutator2}
e^{t^2 \lqu \hat B,\hat A \rqu} &=  \lt e^{i \frac{t}{n} \hat B}  e^{i \frac{t}{n} \hat A} e^{-i\frac{t}{n} \hat B}  e^{-i \frac{t}{n} \hat A} +  O \lt \frac{t^3}{n^3}, \hat A, \hat B\rt \rt^{n^2} \notag \\
&= \lt e^{i \frac{t}{n} \hat B}  e^{i \frac{t}{n} \hat A} e^{-i \frac{t}{n} \hat B}  e^{-i\frac{t}{n} \hat A} \rt^{n^2} + O \lt \frac{t^3}{n}, \hat A, \hat B\rt.
\end{align}
The standard equation for the nested commutator is then derived by replacing in Eq.(\ref{eq:commutator1}) $i t A \rightarrow t^2  \lqu \hat B,\hat A \rqu$. We then obtain
\beq
e^{i t^3 \lqu \hat B,  \lqu \hat B,\hat A \rqu \rqu} = e^{i t \hat B}  e^{t^2 \lqu \hat B,\hat A \rqu} e^{-i t \hat B}  e^{-t^2 \lqu \hat B,\hat A \rqu} + O(t^4, \hat A, \hat B).
\eeq
Similarly as done before, we now use the identity $e^{i t^3 \lqu \hat B,  \lqu \hat B,\hat A \rqu \rqu}  = e^{i \lt \frac{t}{k} \rt^3 \lqu \hat B,  \lqu \hat B,\hat A \rqu \rqu k^3}$ so that we obtain
\begin{align}
\label{eq:intermediate-step}
e^{i t^3 \lqu \hat B,  \lqu \hat B,\hat A \rqu \rqu} &=  \lt e^{i \frac{t}{k} \hat B}  e^{\lt \frac{t}{k} \rt^2 \lqu \hat B,\hat A \rqu} e^{-i \frac{t}{k} \hat B}   e^{-\lt \frac{t}{k} \rt^2 \lqu \hat B,\hat A \rqu} \rt^{k^3} \notag  \\
&+ O \lt \frac{t^4}{k}, \hat A, \hat B\rt.
\end{align}
Now finally we replace Eq.(\ref{eq:commutator2}) in Eq.(\ref{eq:intermediate-step}), but with $t \rightarrow t/k$, i.e. we use $e^{\lt \frac{t}{k}\rt^2 \lqu \hat B, \hat A  \rqu}  = e^{\lt \frac{t}{k l}\rt^2 \lqu \hat B, \hat A  \rqu l^2}$, yielding
\begin{align}
\hspace{-0.6cm}
e^{i t^3  \lqu \hat B, \lqu \hat B,\hat A \rqu \rqu} \hspace{-0.05cm} &= \lt e^{i \hat B \frac{t}{k}} \lt e^{i \hat A \frac{t}{k l }} e^{i \hat B \frac{t}{k l }} e^{-i \hat A \frac{t}{k l }} e^{-i \hat B\frac{t}{k l }}    \rt^{l^2} \right. \notag \\
& \left. e^{-i \hat B \frac{t}{k}}  \lt e^{-i \hat A \frac{t}{k l }} e^{-i \hat B \frac{t}{k l }} e^{i \hat A \frac{t}{k l }} e^{i \hat B\frac{t}{k l }}    \rt^{l^2}  \rt^{k^3} \notag \\
& + O \lt \frac{t^{4}}{k} \rt + O \lt \frac{t^3}{l} \rt.
\end{align}
By using the substitution $i t \hat A \rightarrow t^2  \lqu \hat C, \hat A  \rqu$ it is also possible analogously to derive
\begin{align}
\label{eq:intermediate-stepbis}
e^{i t^3 \lqu \hat B,  \lqu \hat C,\hat A \rqu \rqu} &=  \lt e^{i \frac{t}{k} \hat B}  e^{\lt \frac{t}{k} \rt^2 \lqu \hat C,\hat A \rqu} e^{-i \frac{t}{k} \hat B}   e^{-\lt \frac{t}{k} \rt^2 \lqu \hat C,\hat A \rqu} \rt^{k^3} \notag\\
&+ O \lt \frac{t^4}{k}, \hat A, \hat B, C\rt
\end{align}
and finally
\begin{align}
\hspace{-0.6cm}
e^{i t^3  \lqu \hat B, \lqu \hat C,\hat A \rqu \rqu} \hspace{-0.05cm} &= \lt e^{i \hat B \frac{t}{k}} \lt e^{i \hat A \frac{t}{k l }} e^{i \hat C \frac{t}{k l }} e^{-i \hat A \frac{t}{k l }} e^{-i \hat C\frac{t}{k l }}    \rt^{l^2} \right. \notag \\
& \left. e^{-i \hat B \frac{t}{k}}  \lt e^{-i \hat A \frac{t}{k l }} e^{-i \hat C \frac{t}{k l }} e^{i \hat A \frac{t}{k l }} e^{i \hat C\frac{t}{k l }}    \rt^{l^2}  \rt^{k^3} \notag \\
& + O \lt \frac{t^{4}}{k} \rt + O \lt \frac{t^3}{l} \rt,
\end{align}
which we will need in the following. Setting $\tau = t^3$ we finally obtain Eq.(\ref{eq:nested}).

\section{Rough estimate of the number of operations needed to implement the cross Kerr evolution}
\label{estimate-number-gates}

We estimate the order of magnitude of the number of operations that are needed to implement the cross-Kerr evolution given a desired precision $y$. To provide an estimate, let us first focus on the term $e^{\frac{i \tau \hat p_1^2 \hat p_2^2}{4}}$. Neglecting the Fourier-transforms required to implement $\hat p_1^3$ and $\hat p_2^3$ from the corresponding position-diagonal operators, we obtain that in order to implement the nested commutator in Eq.(\ref{eq:p1p2}) according to Eq.(\ref{eq:nested}) we need
\beq
\label{number-of-operations}
 N_{\tau \hat p_1^2 \hat p_2^2 /4} = [2(4 l^2 + 1) ]  k^3 
  \sim y^{-\frac{9}{2}} \frac{\pi^{\frac{9}{2}}}{18 \cdot 4^4} 
\eeq
elementary operations, where we have only kept the dominant term in $1/y$.
From Eq.(\ref{eq:second-order-splitting}), we need $9$ of these kind of gates, with similar decompositions to $e^{\frac{i \tau \hat p_1^2 \hat p_2^2}{4}}$, in order to implement the operator $\left( e^{i \frac{\pi}{p} \hat n_1 \hat n_2} \rt^p$. Finally, from Eq.(\ref{eq:amplification}) we see that we need to repeat $p$ times these gates. Therefore the total number of gates necessary to implement the cross-Kerr evolution is
\beq
 N_{\pi \hat n_1^2 \hat n_2^2} = 9  N_{\tau \hat p_1^2 \hat p_2^2 /4}  p   \sim y^{-5} \frac{\pi^{\frac{3}{2}}}{2 \cdot 4^4},
\eeq
which results in a thousand gates for $y = 0.1$. 


\section{Decomposition of squeezing and beam splitter in elementary gates}
\label{app:decomposition-gates-exact}

For this analysis, we find it simpler to use the symplectic formalism for the elementary gates as introduced in ~\cite{Gu2009}. In this framework, the relevant gates are represented in terms of their action on the quadrature operators as (using the convention $\Delta^2 q_0 = 1/2$ for the vacuum fluctuations) 
\begin{align}
\label{eq:symplectic}
& \mbox{Squeezing } e^{-i \frac{\ln{s}}{2} (\hat q \hat p + \hat p \hat q )} \rightarrow
\left(\begin{array}{cc} 
\hat q' \\
\hat p'
  \end{array}\right) = 
\left(\begin{array}{cc} 
s & 0\\
 0 & 1/s
  \end{array}\right)
  \left(\begin{array}{cc} 
\hat q \\
\hat p
  \end{array}\right) \notag \\
& \mbox{Rotation } e^{i \frac{\theta}{2} (\hat q^2 + \hat p^2)  } \rightarrow
\left(\begin{array}{cc} 
\hat q' \\
\hat p'
  \end{array}\right) = 
\left(\begin{array}{cc} 
\cos \theta & -\sin \theta \\
\sin \theta & \cos \theta
  \end{array}\right)
  \left(\begin{array}{cc} 
\hat q \\
\hat p
  \end{array}\right) \notag \\
& \mbox{Shear }  e^{\frac{i s q^2}{2}} \rightarrow
\left(\begin{array}{cc} 
\hat q' \\
\hat p'
  \end{array}\right) = 
\left(\begin{array}{cc} 
1 & 0\\
s & 1
  \end{array}\right)
  \left(\begin{array}{cc} 
\hat q \\
\hat p
  \end{array}\right) \notag  \\
& \mbox{FT }  e^{\frac{i \pi}{4} (\hat q^2 + \hat p^2)} \rightarrow
 \left(\begin{array}{cc} 
\hat q' \\
\hat p'
  \end{array}\right) = 
\left(\begin{array}{cc} 
0 & -1\\
1 & 0
  \end{array}\right)
  \left(\begin{array}{cc} 
\hat q \\
\hat p
  \end{array}\right)
\end{align}
\subsection{Decomposition of the squeezing operator (exact)}

We start with the squeezing gate.
It is easy to show that 
\begin{align}
 \left(\begin{array}{cc} 
s & 0\\
 0 & \frac{1}{s}
  \end{array}\right) = &
  \left(\begin{array}{cc} 
0 & -1\\
1 & 0
  \end{array}\right)
  \left(\begin{array}{cc} 
1 & 0\\
s_3 & 1
  \end{array}\right)
  \left(\begin{array}{cc} 
0 & -1\\
1 & 0
  \end{array}\right) \notag \\
 \cdot & 
  \left(\begin{array}{cc} 
1 & 0\\
s_2 & 1
  \end{array}\right)
  \left(\begin{array}{cc} 
0 & -1\\
1 & 0
  \end{array}\right)
  \left(\begin{array}{cc} 
1 & 0\\
s_1 & 1
  \end{array}\right)
\end{align}
with $s_1 = s, s_2 = 1/s, s_3 = s$.
Hence using Eq.(\ref{eq:symplectic}) we obtain Eq.(\ref{eq:decomposition-squeezing}) of the main text.

\subsection{Decomposition of the beamsplitter operation (exact)}

In the proposal of Ref.~\cite{Vasconcelos2010}, squeezed cat states undergo a beam splitter transformation that is described by the symplectic matrix
\beq 
\label{eq:beamsplitter-decomposition}
  \left(\begin{array}{cc} 
q_1' \\
q_2'
  \end{array}\right) =
 \frac{1}{\sqrt{2}} \left(\begin{array}{cc} 
1 &1\\
1 &-1
  \end{array}\right)
  \left(\begin{array}{cc} 
q_1\\
q_2
  \end{array}\right).
\eeq
We want to decompose this beam splitter gate in terms of elementary gates. We use a procedure inspired by Ref.~\cite{Ukai2010b}. For the general change of basis
\beq
  \left(\begin{array}{cc} 
q_1' \\
q_2'
  \end{array}\right) =
 \left(\begin{array}{cc} 
\sqrt{R} &\sqrt{1-R} \\
\sqrt{1-R} & -\sqrt{R}
  \end{array}\right)
  \left(\begin{array}{cc} 
q_1\\
q_2
  \end{array}\right) \equiv M_R   \left(\begin{array}{cc} 
q_1\\
q_2
  \end{array}\right) 
\eeq
(where the obvious identification $R = 1/2$ allows to retrieve the case of interest of the aforementioned beamsplitter) one has the identity
\beq
 \left(\begin{array}{cc} 
M_R & 0 \\
0 & M_R
  \end{array}\right) = 
    \left[  \left(\begin{array}{cc} 
0 & - I_2\\
I_2 & 0
  \end{array}\right) . \left(\begin{array}{cc} 
I_2 & 0 \\
M_R & I_2
  \end{array}\right) \right]^3.
\eeq
This allows us to decompose the beam splitter into Fourier Transform and a gate corresponding to $O(\hat q) = e^{i  (b_1 \hat q_1^2 + b_2 \hat q_2^2 + b_3 \hat q_1 \hat q_2 )}$. To discover the correspondence between the coefficients $b_1,b_2, b_3$ and $R$ we inspect the explicit action of the operator $O(\hat q)$ on the quadratures:
\begin{align}
e^{-i \hat q_1 \hat q_2 b} \hat p_1 e^{i \hat q_1 \hat q_2 b}  =  \hat p_1 +  b \hat q_2  \notag \\
e^{-i \hat q^2  g} \hat p e^{i \hat q^2  g}  =  \hat p +  2 g \hat q
\end{align}
from which it is easy to derive that the operator $O(\hat q)$ leads to
\begin{align}
 \hat p_1 \rightarrow \hat p_1 +  2 b_1 \hat q_1 + b_3 \hat q_2 \notag \\
 \hat p_2 \rightarrow \hat p_2 +  2 b_2 \hat q_2 + b_3 \hat q_1.
\end{align}
This corresponds to the action
\beq
  \left(\begin{array}{cc} 
\vec{q}' \\
\vec{p}'
  \end{array}\right) =
 \left(\begin{array}{cc} 
I_2 & 0 \\
M_R & I_2
  \end{array}\right)   
  \left(\begin{array}{cc} 
\vec{q}' \\
\vec{p}'
  \end{array}\right)
\eeq
with 
\beq
M_R = 
 \left(\begin{array}{cc} 
2 b_1 & b_3 \\
b_3 & 2 b_2
  \end{array}\right).
\eeq
We have hence the identification $b_1 = \sqrt{R}/2, b_2 = -\sqrt{R}/2, b_3 = \sqrt{1 - R}$.
Hence we finally obtain the decomposition for the beam splitter Eq.(\ref{eq:decomposition-bs}).

\subsection{Decomposition of the rotation (exact)}

For completeness, even though we do not use its expression, we also provide a decomposition of the rotation gate.
Due to the identity
\begin{align}
 \left(\begin{array}{cc} 
\cos \theta & -\sin \theta \\
\sin \theta & \cos \theta
  \end{array}\right) = 
 & \left(\begin{array}{cc} 
0 & -1\\
1 & 0
  \end{array}\right)
  \left(\begin{array}{cc} 
1 & 0\\
s_3 & 1
  \end{array}\right)
  \left(\begin{array}{cc} 
0 & -1\\
1 & 0
  \end{array}\right) \notag \\
\cdot &  \left(\begin{array}{cc} 
1 & 0\\
s_2 & 1
  \end{array}\right)
  \left(\begin{array}{cc} 
0 & -1\\
1 & 0
  \end{array}\right)
  \left(\begin{array}{cc} 
1 & 0\\
s_1 & 1
  \end{array}\right)
\end{align}
with $s_1 = \sec{\theta} +  \tan{\theta}  , s_2 = \cos{\theta} , s_3 =   \cos{\theta}  + (1 +  \sin{\theta} )  \tan{\theta} $, using Eq.(\ref{eq:symplectic}) 
we conclude that
\beq
\label{eq:rotation}
e^{i \frac{\theta}{2} (\hat q^2 + \hat p^2)}  = \hat F e^{\frac{i s_3 \hat q^2}{2}} \hat F e^{\frac{i s_2 \hat q^2}{2 }} \hat F e^{ \frac{i s_1 \hat q^2}{2}}.
\eeq

\bibliographystyle{apsrev}
\bibliography{bibliography}

\end{document}